\documentclass[12pt]{article}
\usepackage{amssymb,amsmath}
\usepackage{bm} 
\usepackage{booktabs} 
\usepackage{array}
\usepackage{latexsym}
\usepackage{graphicx}
\usepackage{color}
\usepackage{slashed}
\usepackage{datetime}
\usepackage[nosort]{cite}
\usepackage{verbatim}
\usepackage{enumerate}
\usepackage{makecell} 
\usepackage{chngpage} 
\allowdisplaybreaks
\usepackage{psfrag}
\usepackage{mathtools}
\usepackage{mciteplus}
\usepackage{dsfont}
\usepackage{multirow} 
\usepackage{arydshln} 

\usepackage[colorlinks=true,      linkcolor=blue,      urlcolor=blue,      
            filecolor=blue,      citecolor=blue,       pdfstartview=FitH,     
						pdfpagemode=UseNone,      bookmarksopen=true]{hyperref}  
\usepackage[all]{hypcap}     

\newcommand{\comm}[1]{} 

\def\({\left(}
\def\){\right)}
\def\[{\left[}
\def\]{\right]}

\def\One{{\hbox{ 1\kern-.8mm l}}}

\def\barray{\begin{array}}
\def\earray{\end{array}}
\def\be{\begin{equation}}
\def\ee{\end{equation}}
\def\bea{\begin{eqnarray}}
\def\eea{\end{eqnarray}}
\def\bal{\begin{align}}
\def\eal{\end{align}}

\numberwithin{equation}{section} 
\allowdisplaybreaks  

\makeatletter
\g@addto@macro\bfseries{\boldmath}
\makeatother

\definecolor{cardinal}{rgb}{0.6,0,0}
\definecolor{darkgreen}{rgb}{0,0.4,0}
\definecolor{golden}{rgb}{0.92, 0.7, 0}
\definecolor{midnight}{rgb}{0, 0, 0.5}
\definecolor{darkblue}{rgb}{0, 0, 0.7}
\definecolor{purple}{rgb}{0.5, 0, 0.5}

\def\IR{\mathbb{R}}

\def\cF{{\cal F}}

\def\nBPS#1{$\frac{1}{#1}$-BPS}

\def\cO{{\cal O}}

\def\gh{{\gamma^{(1)}}}
\def\ghh{{\gamma^{(2)}}}

\usepackage{epsf}
\usepackage{cite}
\usepackage{fancyhdr}
\usepackage{bm}
\usepackage{enumerate} 

\usepackage{empheq} \empheqset{box=\fbox}
\usepackage[usenames,dvipsnames]{xcolor}

\numberwithin{equation}{section}  
\allowdisplaybreaks

\usepackage{graphicx}
\usepackage{tikz}
\usetikzlibrary{decorations.markings}
\tikzset{->-/.style={decoration={
			markings,
			mark=at position #1 with {\arrow{stealth}}},postaction={decorate}}}
\usepackage{adjustbox}
\usepackage{pgfplots}
\pgfplotsset{compat=1.11}
\usepgfplotslibrary{fillbetween}
\usetikzlibrary{intersections}
\usetikzlibrary{patterns}

\pgfdeclarelayer{bg}
\pgfsetlayers{bg,main}

\usetikzlibrary{calc}
\tikzset{
	samples=100,
}
\pgfplotsset{compat=1.11}

\pgfmathsetmacro\T{3.14}
\pgfmathsetmacro\A{0.2}
\pgfmathsetmacro\N{4}
\pgfmathsetmacro\D{\N*\T}

\begin{document}


\begin{flushright}

\end{flushright}

\vspace{3mm}

\begin{center}

{\huge {\bf Resolving Black-Hole Microstructure \\}}
\vspace{3mm}
{\huge {\bf with New Momentum Carriers}}

\vspace{14mm}

{\large
\textsc{Iosif Bena$^a$, Nejc \v{C}eplak$^a$, Shaun Hampton$^a$, Yixuan Li$^a$, Dimitrios Toulikas$^a$, Nicholas P. Warner$^{abc}$}}
\vspace{12mm}

\textit{$^a$Universit\'e Paris Saclay, CNRS, CEA,\\
Institut de Physique Th\'eorique,\\
91191, Gif-sur-Yvette, France
\\[12 pt]
\centerline{$^b$Department of Physics and Astronomy}
\centerline{and $^c$Department of Mathematics,}
\centerline{University of Southern California,} 
\centerline{Los Angeles, CA 90089, USA}}

\medskip

\vspace{4mm} 
%

{\footnotesize\upshape\ttfamily  iosif.bena, nejc.ceplak, shaun.hampton, yixuan.li, dimitrios.toulikas  @ ipht.fr; warner @ usc.edu.} \\
\vspace{13mm}

\textsc{Abstract}

\end{center}

\begin{adjustwidth}{10mm}{10mm} 
 
\vspace{1mm}
\noindent

All known horizonless black-hole microstate geometries correspond to brane sources that acquire a finite size, and hence break the spherical symmetry of the black hole. 
We construct, for the first time, solutions with zero horizon area that have the same charges as a  three-charge F1-NS5-P Type-IIA black hole and preserve this spherical symmetry. The momentum of these solutions is carried by longitudinal D0-D4 density fluctuations inside the NS5-branes.
We argue that these solutions should be interpreted as the long-throat limit of a family of smooth, horizonless microstate geometries, called superstrata, where such geometries degenerate.
The existence of these geometries indicates that a finite-size horizon does not appear even in the singular corners of the moduli space of three-charge microstate geometries.

\end{adjustwidth}

\thispagestyle{empty}
\clearpage



\baselineskip=14.5pt
\parskip=3pt

\tableofcontents

\baselineskip=15pt
\parskip=3pt


\section{Introduction}
\label{sec:Intro}

One of the remarkable achievements of string theory is that it can provide a microscopic description of  black-hole entropy.  
It was found that, at vanishing string coupling, different string/brane configurations  could reproduce the  Bekenstein-Hawking entropy of the corresponding  black hole  
\cite{Sen:1995in,Strominger:1996sh}.   The black-hole geometry, and its horizon, then emerge as the string coupling, and hence Newton's constant, $G_N$, becomes finite.   Indeed, the horizon grows with $G_N$ \cite{Horowitz:1996nw, Damour:1999aw, Chen:2021dsw}, but because  gravity generically compresses matter, it was believed that all the perturbative string states would  collapse behind a horizon.   Thus the perturbative microstates,  whose counting gives the black-hole entropy, would not be visible once gravity takes effect.

Insights from brane physics show that this picture is too na\"ive.
The tension of  D-branes and NS-branes decreases as the coupling increases, and so adding momentum excitations causes them to spread in directions transverse to their world-volume.  Indeed, it was noted in  \cite{Bena:2004wt} that three-charge brane configurations carrying momentum would grow with $G_N$ at the same rate as the black-hole horizon.  It was then found that three-charge horizonless geometries supported by topological fluxes have the same behavior \cite{Bena:2005va, Berglund:2005vb,Gibbons:2013tqa}. Thus was born the Microstate Geometry (MG)  Programme in which one constructs {\it smooth, horizonless geometries} that approximate the classical black-hole solution everywhere except at the  horizon scale, where MG's end in a smooth, horizonless cap. 

Microstate Geometries are part of a  larger framework, known as the \emph{Fuzzball Programme}. The defining ideal of this programme is that individual black-hole microstates, generically referred to as fuzzballs,  must be horizonless because horizons imply entropy and give rise to information loss \cite{Mathur:2008nj, Almheiri:2012rt}.
Fuzzballs have the same mass, charge and angular momentum as a given black hole and can be arbitrarily quantum and arbitrarily strongly curved.  
They describe pure states of the black hole and, if a holographic description is available, are dual to  pure states of the CFT that can be used to account for the black-hole entropy. 
Microstate Geometries fit in this paradigm as the string-theory fuzzballs that are sufficiently coherent as to become well approximated by smooth solutions of supergravity.

There also exist fuzzballs that are not smooth supergravity solutions but can be described using other well-defined limits of string theory.  Indeed, this led to the definition of a {\it Microstate Solution},  \cite{Bena:2013dka}, which is a  horizonless solution of supergravity, or a horizonless, physical limit of a supergravity solution,  that has the same mass, charge and angular momentum as a given black hole.  Microstate solutions are allowed to have singularities that either correspond to brane  sources, or can be patch-wise dualized into a smooth solution.   In this paper we will  refine  this classification further to distinguish microstate solutions  corresponding to  pure states from  {\it Degenerate Microstate Solutions}, which correspond to a limited  family of microstates.

It is important to emphasize that Fuzzballs are all, by definition, horizonless, regardless of whether they can be described within supergravity.
In this paradigm, horizons arise only as a consequence of averaging over microstates and are thus necessarily related to ensembles of such states.   This is what leads to the entropy-area relation. 
But if pure states correspond to horizonless microstates, then a solution with a horizon should not describe the physics of {\it any} pure state of the system and should not be holographically related to any pure state of the dual CFT.\footnote{This has only been shown so far for (0+1)-dimensional CFT's dual to asymptotically-AdS$_2$ spacetimes \cite{Bena:2018bbd}.}

The purpose of this paper is to make some steps towards the resolution 
of what  appears to be a counterexample to the Fuzzball paradigm: the possibility that some pure CFT states are dual to a supergravity solution with a horizon.  The putative counterexample comes from a singular limit of a class of Microstate Geometries known as {\it superstrata.}

Superstrata are horizonless solutions that have the same charges as a D1-D5-P supersymmetric black hole. They are, perhaps,  the most analyzed and well-studied of all MG's \cite{Bena:2015bea,Bena:2016ypk,Bena:2017geu,Bena:2017xbt,Bena:2018mpb,Ceplak:2018pws,Bena:2019azk,Heidmann:2019zws,Heidmann:2019xrd,Mayerson:2020tcl, Shigemori:2020yuo, Bena:2020yii,Bena:2020iyw,Giusto:2020mup,Martinec:2020cml,Houppe:2020oqp,Ceplak:2021wak, Ceplak:2021kgl,Ganchev:2021pgs, Ganchev:2021iwy},  and the holographic dictionary for these geometries is  now well-established  \cite{Kanitscheider:2006zf,Kanitscheider:2007wq,Taylor:2007hs,Giusto:2015dfa,Bombini:2017sge,Giusto:2019qig, Tormo:2019yus, Rawash:2021pik, Ganchev:2021ewa}.  
The corresponding black holes have an infinitely-long AdS$_2$ throat, but in superstrata, this throat is capped off at a large but finite depth, which is inversely proportional to a parameter, $a$, that controls the angular momentum, and  the spatial extent of the configuration.  The momentum charge of a superstratum is  carried by  flux excitations whose Fourier amplitudes give an additional set of parameters, $b_n$.  The problematic limit, and putative counterexample, arises as one takes $a \to0$.

These parameters have a well-understood interpretation in the dual D1-D5 CFT  \cite{Giusto:2019qig}. The CFT states dual to superstrata are constructed starting from  RR-ground states that are usually described as having  $(+,+)$ strands and $(0,0)$ strands.\footnote{For explanation of this notation, see, for example, \cite{Avery:2010qw,Bena:2015bea}.} The former carry  angular momentum but  no momentum, and their number is proportional to $a^2$.  The $(0,0)$ strands have vanishing angular momentum but, in the superstratum, carry momentum excitations with a quantum number, $n$.  The number of such excited strands is proportional to $b_n^2$ and the total momentum charge is given by:
\begin{align}
Q_P  ~\sim~   \sum_{n=1}^\infty \, n\, b_n^2\,.
	\label{eq:Mom}
\end{align}
Requiring the superstrata to be smooth and free of closed time-like curves imposes a constraint of the schematic form:   
\begin{align}
	\frac{Q_1 Q_5}{R_y^2}  ~=~  a^2 ~+~  \frac{1}{2}\,  \sum_{n=1}^\infty \, b_n^2\,,
	\label{eq:reg10n}
\end{align}
where $Q_1$ and $Q_5$ are the supergravity D1 and D5-brane charges and $R_y$ is the asymptotic radius of the common D1-D5 direction.
The important point is that adding more momentum-carrying modes (by increasing the $b_n$'s) makes  $a$ smaller, so the AdS$_2$ throat becomes longer, capping off at higher and higher red-shifts.  In the $a \to 0$ limit, the cap moves to infinite redshift and the superstratum solution appears to become identical to the classical extremal D1-D5-P black hole.

From the perspective of the dictionary to the dual CFT, this limit appears well-defined and corresponds to
a pure state with only $(0,0)$ strands.
Thus it appears that as one moves in the space of CFT states dual to superstrata, one encounters some pure states whose bulk dual has a horizon.  This  violates the basic principle of the Fuzzball/MG programme: Pure states should not be dual to a configuration that has a horizon. 

As we discuss in Section \ref{sec:MCarriers}, the appearance of a horizon is explained by noting that
in the D1-D5-P frame, the standard superstratum construction not only restricts the momentum-carrying excitations, but also involves a smearing operation.
This smearing preserves the details of the microstructure only when $a\neq 0$, while  in the $a \to 0$ limit it  averages over  distinct momentum-carrying configurations and this gives rise to a solution with a horizon. 
If one avoids this smearing, and takes into account the degrees of freedom this smearing erases, the geometry  remains horizonless even  as $a \to 0$.

In this paper we show how this can be achieved by constructing a new class of three-charge solutions with vanishing horizon area that go beyond the standard superstratum construction by incorporating additional momentum-carrying excitations.
We do this by working in the Type IIA F1-NS5-P duality frame, and the new momentum carriers that can resolve the microstructure are  D0-brane and D4-brane charge densities that vary along the common F1-NS5 direction.
These excitations have the important property that, unlike all other microstate geometries, they carry momentum \emph{without} expanding the branes in directions transverse to their world-volume. Hence, one can think of them as giving rise to a longitudinally polarized momentum wave on branes that remain localized at a single point in the transverse directions, and do not break the rotational $SO(4)$ symmetry of the black-hole solution. 

Since duality transformations preserve degrees of freedom while encoding them in different ways, our  Type IIA F1-NS5-P  supergravity solutions must have counterparts in the D1-D5-P frame.  However, to get from one frame to the other, one must perform a T-duality along the common F1-NS5 direction, and the solutions we construct depend explicitly on this direction.  As a result, our Type IIA supergravity solutions  become configurations involving a coherent set of higher Kaluza-Klein modes, and thus cannot be described as D1-D5-P solutions in Type IIB supergravity.\footnote{It is also interesting to note that the exact same phenomenon happens when one tries to dualize D1-D5-P superstrata that depend on the common D1-D5 direction to the IIA F1-NS5-P duality frame we consider: the smooth geometries are dualized into microstate solutions that contain excited towers of KK modes and are not describable in supergravity.}

The main result of this paper is the solution given in  equation~\eqref{eq:NS5F1P-D0D4}: It  represents a family of three-charge F1-NS5-P solutions with D0 and D4 densities and no macroscopic horizon.
Globally, this solution preserves the four supercharges of the corresponding three-charge black hole. 
However, if we zoom in at a fixed location along the  F1 and NS5-branes, we find that the configuration {\it locally} preserves eight supercharges.  In this limit, the local D0 and D4 densities are approximately constant and the solution preserves eight Killing spinors, four of which are identical to those of the F1-NS5-P black hole. Hence, near the brane sources the solution behaves locally like a two-charge system with a vanishing horizon area.

It is important to emphasize that the solution presented here is a singular brane configuration with vanishing horizon area, and its role as a fuzzball needs clarification. 
As originally conceived, a {\it Microstate Solution} is a horizonless, but singular, brane configuration that corresponds to a   black-hole microstate that can be fully resolved in string theory.  We need to broaden this idea to include {\it Degenerate Microstate Solutions}.  Such an object is defined to be a singular supergravity solution with the following properties:
\begin{itemize}
 \setlength\itemsep{-2pt}
	\item It must have vanishing horizon area.  
	\item  The source must correspond to a well-defined family of branes. 
	\item   The microstructure of the brane source can be revealed, and counted, through standard string theory methods.
	 \item There must be geometric deformations, or transitions, that can  resolve the solution into microstate solutions or microstate geometries.
\end{itemize}

One of the features of microstate solutions, and microstate geometries, is that if one zooms into their cores, the underlying geometric elements are ``locally primitive,'' which means that they locally preserve 16 supercharges. Taken as a whole, the complete solution preserves only a subset of these supercharges. By contrast, the cores of degenerate microstate solutions will typically preserve only 8 supercharges.  This is too much supersymmetry for the configuration to generate a  horizon, and so the underlying structure can still be accessed and probed by string theory.  However, the reduction from 16 supercharges to 8 supercharges reflects the fact that such solutions still correspond to a family of individual microstates, but this family is  too small to generate a horizon in supergravity.

 In the past, the configurations we are classifying as degenerate microstate solutions have  sometimes been said to have ``small'' (string-scale) horizons because they represent stringy ensembles of states.  We prefer the  defining ideas of degenerate microstate solutions because they accentuate  the accessibility of the microstructure to stringy analysis and geometric resolutions, while the cloaking of such things in horizons is, once again, just code for ensemble averaging of microstructure. 
 
The archetype of a  degenerate microstate solution is, of course, the pure D1-D5 solution, whose microstructure has been throughly understood in string theory \cite{Lunin:2001fv,Lunin:2002iz,Chen:2014loa}.  As we will discuss, the degenerate microstate solutions that we will construct in this paper are, at their core, equivalent to D1-D5 degenerate microstate solutions. In subsequent work we plan to explore geometric transitions  that will  resolve  these degenerate microstate solutions into microstate solutions and microstate geometries. 
In Section~\ref{sec:MCarriers} we describe the general features of the standard superstratum construction and how it neglects some degrees of freedom and necessarily results in smearing in the $a \to 0$ limit.
We also discuss the supersymmetries preserved by the solution.
In Section~\ref{sec:Construction} we describe the construction of the eight-supercharge NS5 solution with D0-D4 charges that carry momentum without transverse fluctuations. We then add coherent F1-string excitations to this system, and obtain the complete supergravity description.  It is  this microstate solution that provides  the resolution of the  $a\to0$  limit: a solution with black-hole charges, vanishing horizon area, and SO(4) symmetry. 

In Section~\ref{sec:Analysis} we analyze this new geometry and compare it to the three-charge black-hole solution.   Section~\ref{sec:Discussion} contains a discussion of our results and an outline  of possible future research.
Some of the details of the construction that are omitted in Section~\ref{sec:Construction} are presented in Appendix~\ref{app:Dualitites}. In Appendix~\ref{app:Conventions} we collect some of the conventions used throughout the paper.

\section{Momentum carriers on superstrata}
\label{sec:MCarriers}

In five dimensions, a BPS black hole only has a finite-sized horizon if it has three charges and thus preserves four supercharges (\nBPS{8}).  The corresponding  microstate geometries and microstate solutions must globally preserve the same supercharges, however  their {\it cores} can have more supersymmetries {\it locally.}  Indeed, their fundamental  building blocks are locally {\it primitive} and have 16 supercharges  \cite{Bena:2011uw}, but have fewer supersymmetries when considered globally because their shapes and dipolar charge distributions break the supercharges down to the universal subset that is common to the entire configuration.  

Since microstate geometries and microstate solutions are supported by sources  that have locally more supersymmetries than the black hole, they do not have in general an event horizon. 
Indeed, the existence of superstrata was originally conjectured based on a double-bubbled geometric transition of the D1-D5 system  \cite{Bena:2011uw}.  Specifically, if one starts with  a stack of  D1-branes and adds a momentum wave, then the configuration is globally \nBPS{4} but locally  \nBPS{2}.  If one then combines a D1-brane with a profile carrying a momentum wave with a D5-brane with the same profile, the system  is globally \nBPS{8} but locally  \nBPS{4}.  By adding angular momentum and a KKM dipole charge, one can make a geometric transition to a momentum carrying object that is  globally \nBPS{8} but locally  \nBPS{2}. The result is a superstratum \cite{Bena:2015bea}.

To make a smooth geometry, the ``special direction'' of the KKM must coincide with the common D1-D5 direction, which we  parameterize with the coordinate $v$.%
\footnote{To be more precise, the common D1-D5 direction is described by a periodic coordinate $y$, while $v$ is a null coordinate: see equation \eqref{eq:NullCC}. Supersymmetry requires the solution to be independent of the other null coordinate, $u$, and  one can think of the latter as describing ``time''  while $v$ denotes the ``spatial'' coordinate (see also \cite{Bena:2017geu} for a more careful discussion).}
However,  the standard Kaluza-Klein Monopole (KKM) geometry must be $v$-independent,\footnote{One can obviate this difficulty by allowing higher Kaluza-Klein modes in the monopole, but this takes us outside of Type-IIB supergravity.} and this conflicts with the addition of momentum excitations, which necessarily depend on $v$.   Indeed, the $v$-circle pinches off at the KKM location, and so one cannot source $v$-dependent fluctuations on the KKM locus without creating a singularity.

This difficulty was resolved in  \cite{Niehoff:2013kia} and is best understood by starting from the standard, maximally-spinning supertube \cite{Mateos:2001qs,Emparan:2001ux}.  One takes the D1-D5 system and adds a KKM dipole and angular momentum so that the supertube wraps a circle in an $\IR^2$ of the $\IR^4$ transverse space.  The angle along this circle is denoted by $\phi$, and the solution is independent of $(\phi,v)$.   This describes the maximally-spinning \nBPS{4} supertube and it corresponds to a coherent superposition of RR ground states in the CFT consisting of only $(+,+)$ strands.   One can now allow the density of D1- and D5-branes to vary {\it along the $\phi$ direction of the supertube}.
In terms of the standard mode numbers inherent in superstrata, $(k,m,n)$, this density fluctuation corresponds to a  $(k,0,0)$  excitation.	
The result is still a \nBPS{4} supertube, and it is still $v$-independent, but it is now a mixture of $(+,+)$  and  $(0,0)$ strands (of length $k$).  The numbers of such strands is determined  by Fourier  coefficients, $a$ and $b_{k,m=0,n=0}$.  
 
In superstrata one can think of the $(0,0)$ strands (or the $\phi$-dependent density fluctuations in the $(k,0,0)$ solution) as the ``medium'' that carries the momentum, and the solutions where these modes are excited have generic values of $(k,m,n)$. One necessarily has $k >0$ because the momentum is being carried by the density fluctuations around $\phi$.  As discussed in detail in  \cite{Niehoff:2013kia}, the $v$-dependent fluctuations are not, and cannot be, sourced on the original supertube locus: these fluctuations are delocalized in the fluxes and geometry of the topologically-non-trivial three-cycles of the D1-D5-KKM solution.

The $a \to 0$ limit of superstrata is motivated by the desire to construct  solutions with vanishing angular momentum that resemble a black hole with arbitrary precision.\footnote{As explained in \cite{Bena:2018bbd}, there are two such limits. In the first limit, one keeps finite the energy of asymptotic observers and the asymptotic structure of spacetime, and the AdS$_2$ throat becomes longer and longer and its cap becomes deeper and deeper, approaching the infinite throat of the supersymmetric black hole. In the second limit, one keeps finite the energy of an observer in the cap, and in this limit the cap remains fixed, while the asymptotic structure of the solution becomes AdS$_2$ times a compact space. This discussion is about the first limit.} In view of the previous discussion it is now evident just how pathological  this limit  is for standard superstrata. Namely, by keeping the UV unchanged and taking $a \to 0$, one is collapsing both the supertube that defines the momentum carriers and the topological bubble that supports the momentum-carrying fluxes. The end result is to push the KKM locus and the center of the $\IR^4$ base-space of the solution to a point, while keeping the momentum fixed. Since the KKM forces $v$-independence, the momentum charge only survives in this limit because the momentum carriers are smeared along the $v$-circle and as a result the geometry develops a horizon.
Hence, the standard superstratum momentum carriers, which are $v$-dependent and have polarizations in the $\IR^4$ directions are crushed to a point in the transverse space and smeared along the $v$-direction in the $a \to 0$ limit.

In the dual CFT picture, the $a\to 0$ limit of various superstratum solutions corresponds to various states that only have  $(0,0)$ but no  $(+,+)$ strands, and hence have no angular momentum. Hence, these pure states appear naively to be dual to a bulk solution with a horizon. Furthermore, the bulk information that distinguishes these pure states from one another appears to vanish in this limit. Thus in the limit of vanishing angular momentum, the superstratum holographic dictionary appears to break down.  In order to solve this puzzle, and the apparent loss of information in the holographic dictionary, we need to consider all possible momentum carriers of the system, and, in particular, find the modes that carry momentum and have vanishing angular momentum in the space-time.  The simplest duality frame in which one can build these modes is the Type IIA frame in which the three charges of the black hole correspond to F1 strings, NS5-branes and  momentum.\footnote{It is also possible to add such fluctuations in the D1-D5-P duality frame, but these correspond to fluctuations of brane and string densities that wrap partially the $T^4$ compact space, and hence break the  isotropy of the torus. The advantage of the IIA F1-NS5-P frame is that these modes preserve the $T^4$ isotropy.}

One can relate this frame very easily to the normal IIB D1-D5-P frame by an S-duality to a Type IIB F1-NS5-P system, followed by a T-duality. In this duality frame, the NS5-brane can carry momentum along the common F1-NS5 direction by the excitation of the  internal scalar field of the Type IIA NS5-brane. This corresponds in supergravity to turning on fluctuating Ramond-Ramond fields $C_1$ and $C_3$, that can be thought of as coming from D0- and D4-brane density fluctuations inside the NS5-brane. These density fluctuations can be chosen to integrate to zero, so that the total solution only has F1, NS5 and P charge.
These momentum-carrying excitations have vanishing angular momentum in the transverse $\IR^4$ space and are well-defined even in the $a \to 0$ limit.  

\begin{figure}[t]
	\centering
	\label{fig:MTheoryPic}
		\begin{tikzpicture}
			\begin{scope}[shift={(0,0)}]
				\draw[fill=ForestGreen!8, thick] (0,0) rectangle (4*3.14,4);  
				\draw [black] (0,0) rectangle (4*3.14,4); 
				\draw[domain=0:4*3.14, color=blue, smooth, variable=\x, thick]   plot ({\x},{2.+ 1*sin(2*  \x r)});
				\node[align = center, centered] at (-0.5, 2.) {$x_{10}$};
				\node[align = center, centered] at (2*3.14, -0.5) {$v$};
				\node[align = center, centered, blue] at (3.75*3.14, 0.5) {M5};
				\draw[red, -stealth, thick] (11,2.7)--(12,2.7);
				\node[align = center, centered, red] at (11.5, 3) {$P$};
				\node[align = center, centered, ForestGreen] at (1,4.3 ) {M2};
			\end{scope}
		\end{tikzpicture}
	\caption{Initial configuration in the M-theory frame: M2-branes (green) are wrapping the $S^1(v)\times S^1(x_{10})$ circles, while the M5-branes (blue) wrap the $S^1(v)\times T^4$ ($T^4$ is not pictured) and have a wave carrying a momentum, $P$, along $v$. 
		The M5-branes with a momentum wave have a non-trivial profile in the $S^1(v)\times S^1(x_{10})$ plane, and hence have locally non-zero M5 charges parallel to the $x_{10}$ direction, as well as non-trivial momentum along $x_{10}$. 
		When one compactifies this M-theory solution to Type IIA along $x_{10}$, these charge components  become  D4 and D0 charge densities respectively.}
\end{figure}
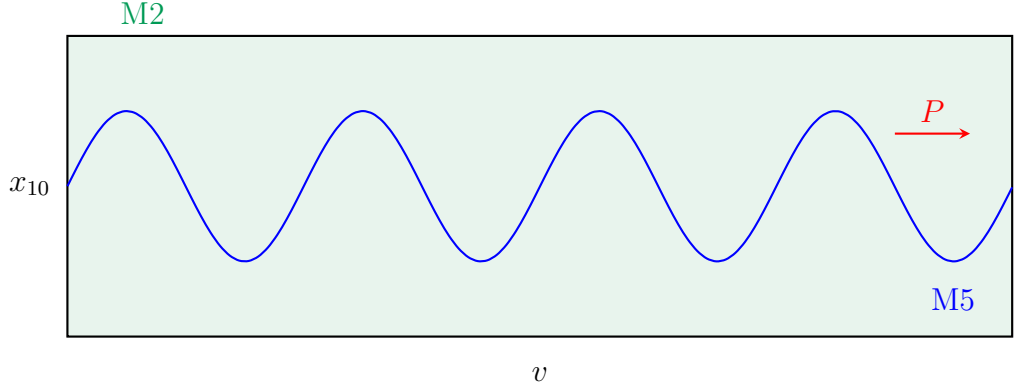
The fact that adding D0-D4 dipole charges to the F1-NS5 system  is natural is perhaps best understood by going to the M-theory frame. 
Consider Type IIA theory on $\IR^{1,4}\times S^1(v)\times T^4$ and denote the M-theory circle by $S^1(x_{10})$.
The  F1-NS5 system  lifts to a configuration of M5  and M2-branes, where the M5-branes wrap  $T^4 \times S^1(v)$ and the M2-branes wrap $S^1(x_{10}) \times  S^1(v)$. 
The D0-D4 densities carry momentum as a longitudinal wave along the common direction in  the F1-NS5 system.  In M-theory the NS5-D0-D4-P subsystem uplifts to a momentum-carrying wave on the M5-brane, whose transverse  polarization is strictly in the M-theory direction. This M5-brane has 8 supersymmetries, but if one zooms near the profile at a specific location one finds an M5-brane with orthogonal momentum, which preservers 16 supercharges. 
When one reduces this configuration along the $x_{10}$ direction to ten-dimensional Type IIA theory, the momentum and M5-charge polarized along the $x_{10}$ become D0 and D4 charge densities.

This leads to the starting point of our analysis:
Our aim is to construct three-charge Type IIA supergravity  solutions with F1-NS5-P charges, where the momentum is carried by fluctuating D0-D4 density waves.  In contrast to all the three-charge horizonless solutions constructed so far, our solutions are $SO(4)$ singlets under rotations on the $\mathbb{R}^4$ base space, exactly as the black hole.  Furthermore, these solutions are \nBPS{8} (4 supercharges) globally, but \nBPS{4} (8 supercharges) locally, and hence have a vanishing horizon area.  But as we explained earlier, the result of our analysis will be a new  family of degenerate microstate solutions.

\section{Construction of the new three-charge solution}
\label{sec:Construction}

Our construction starts from the well-known solution for the F1-P system in Type IIB supergravity in ten dimensions. We then use a series of S-dualities and T-dualities (whose details are presented in Appendix~\ref{app:Dualitites})
to arrive at the geometry corresponding to the two-charge NS5-P system with local D0-D4 charges.
We then add a fundamental string charge to this system. We do this by applying an S-duality and then a T-duality to the initial frame which results in a system with  D5 and P charges.  In that duality frame one can add a D1 charge in a straightforward manner. After we add the D1 charge, reversing the last duality chain takes us to the solution we are seeking: One which carries F1-NS5-P charges, has SO(4) spherical symmetry and vanishing horizon area.

\subsection{Generating an NS5-P solution with local D0-D4 charges}

\subsubsection{Starting point: the F1-P solution with a non-trivial $T^4$ profile}

The solution in $D$ spacetime dimensions sourced by a fundamental string carrying momentum lies entirely in the NS sector of the theory, and is given by \cite{Callan:1995hn, Dabholkar:1995nc}:
\begin{subequations}
	\label{eq:F1PSolGen}
	\begin{align}
		ds^2 &= -\frac2{H} \, dv\left[ du -\frac{\dot F^2(v)}{2}  \left(H- 1\right)\,  dv + \dot F_M(v) \, \left(H- 1\right) \, dx^M\right] + \delta_{MN} dx^M\, dx^N\,,\\
		B &= -\left(1- \frac{1}{H}\right)\,\left[ du \wedge dv + \dot F_M(v)\, dv \wedge dx^M\right]\,,\qquad 
		e^{2 \phi}= \frac{1}{H}\,,
	\end{align}
\end{subequations}
with all other fields vanishing. 
The coordinates $u$ and $v$ define the  light-cone directions along the world-sheet of the string.
The remaining transverse directions are parameterized by Cartesian coordinates, $x_M$, with $M = 1,\ldots D-2$. The shape of the string is given by profile functions, $F_M(v)$, with the dot denoting the derivative with respect to $v$. The string sources a warp factor which is a harmonic function, $H$, in the $D-2$  dimensional transverse space:
\begin{align}
	\label{eq:HarFunGen}
	H \equiv 1 + \frac{Q}{\left|x_M - F_M(v)\right|^{D-4}}\,,
\end{align}
where $Q$ is the supergravity charge associated to the fundamental string and is proportional to the ADM mass per unit length \cite{Dabholkar:1995nc}.

We take the space-time to be ten-dimensional with the topology  $ \IR_t\times \mathbb{R}^{4}\times S^1(y)\times T^4$.  We will refer to the $\mathbb{R}^4$ as the base space, and it will be parameterized by   $x_i$, with $i = 1,2,3,4$, while the  $T^4$ will be parameterized by $z_a$ with $a = 6,7,8,9$. 
We take the radius of the circle $S^1(y)$ to be given by $R_y$, and the coordinate $y$ is periodically identified with $y \sim y + 2\pi\, R_y$. 
The null coordinates appearing in \eqref{eq:F1PSolGen} are related to the usual spacetime coordinates through:%
\footnote{Note that compared to \cite{ Callan:1995hn, Dabholkar:1995nc}, we have rescaled $u$ and $v$ by a factor of $\sqrt{2}$.}
\begin{align}
	\label{eq:NullCC}
	v = \frac{t+y}{\sqrt2}\,, \qquad u = \frac{t-y}{\sqrt2}\,.
\end{align}

We choose the momentum-carrying string to wrap the compact $y$ direction and to be localized at the origin of $\mathbb{R}^4$.  
For simplicity, we  take the string to oscillate along one of the directions of the torus, $z_9$.    Since we are interested in a solution that is isotropic along the torus, we  smear the string source along the full $T^4$. 
The corresponding profile function is
\begin{align}
	\label{eq:Profile1}
	F_M(v) = \delta_{Ma}\,c_a + \delta_{M9}\,F(v)\,,
\end{align}	
where $F(v)$ is an arbitrary periodic function of period $\sqrt{2} \pi R_y$ and we include constants $c_a$ which are integrated over in the process of smearing.
\begin{figure}[t!]
	\centering
	\label{fig:F1P}
		\begin{tikzpicture}
			\begin{scope}
				\clip (0,0) rectangle (4*3.14,4); 
				\foreach \z in {-5,...,50}
				\draw[domain=0:4*3.14, color=blue, smooth, variable=\x, dotted]   plot ({\x},{0.25*\z+ 1*sin(2*  \x r)});
			\end{scope}
			\begin{scope}[shift={(0,0)}]
				\draw [black] (0,0) rectangle (4*3.14,4); 
				\draw[domain=0:4*3.14, color=blue, smooth, variable=\x, thick]   plot ({\x},{2+ 1*sin(2*  \x r)});
				\node[align = center, centered] at (-0.5, 2) {$z_9$};
				\node[align = center, centered] at (2*3.14, -0.5) {$y$};
				\node[align = center, centered, blue] at (13, 2) {F1};
				\draw[red, -stealth, thick] (11,2.7)--(12,2.7);
				\node[align = center, centered, red, fill=white] at (11.5, 3.) {$P$};
				\node[align = center, centered, blue, fill=white] at (7/4*pi, 0.5) {$F(v)$};
				%
			\end{scope}
		\end{tikzpicture}
	\caption{The shape of the fundamental string in 	the $y-z_9$ plane at a fixed time $t$. 
		The string is wrapping the $y$-circle while its profile in the  $z_9$ direction is given by an arbitrary periodic function $F(v)$. 
		The system has a global F1 charge and a global momentum charge, denoted by $P$.
		Finally, the profile is smeared on the  $S(z_9)$ circle, the smearing process being here depicted with the dotted lines.
		The non-trivial profile results in local variations of the charges in the $z_9$ and $y$ directions.  
	}
\end{figure}
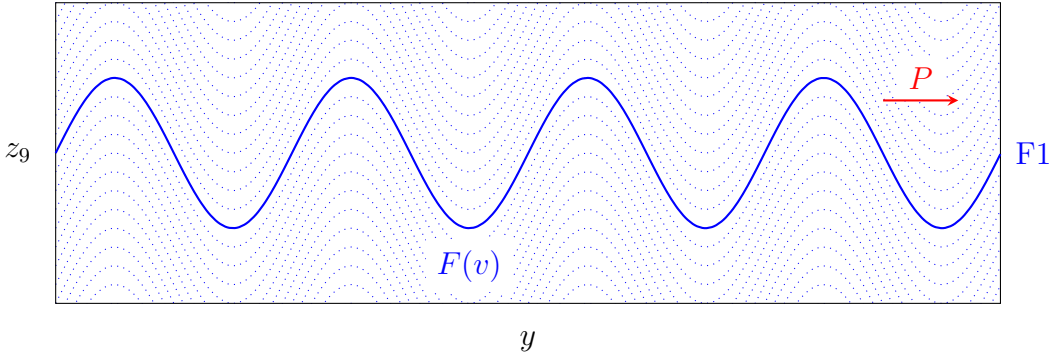
The solution after smearing on the torus (see also Figure~\ref{fig:F1P}) is
\begin{subequations}
	\label{eq:F1PSol}
	\begin{align}
		ds^2 &= -\frac2{H_5} \, dv\left[ du -\frac{\dot F^2(v)}{2}  \left(H_5- 1\right) \, dv + \dot F(v) \, \left(H_5- 1\right) \, dz^9\right] +  dx^i\, dx^i+ dz^a\, dz^a\,,\\*
		B &= -\left(1- \frac{1}{H_5}\right)\,\left[ du \wedge dv + \dot F(v)\, dv \wedge dz^9\right]\,,\qquad 
		e^{2 \phi}  = \frac{1}{H_5}\,,
	\end{align}
\end{subequations}
where  the harmonic function \eqref{eq:HarFunGen} is now given by%
\footnote{The label is added to the harmonic function and to the charge for future convenience.}
\begin{align}
	\label{eq:HarFun}
	H_5(r) = 1 + \frac{Q_5}{r^2}\,,\qquad r^2 = x_i \, x_i\,.
\end{align}
The profile of the momentum-carrying wave, $F(v)$, is arbitrary in the $y-z_9$ plane, so the system has locally varying F1 and momentum charge densities, which generically source the metric and $B$-fields with   components both along the $y$-direction and along the  $z_9$-direction.  We denote these configurations as F1$(y)$,  P$(y)$,  and  F1$(z_9)$, P$(z_9)$, respectively (see figure~\ref{fig:LocalFig}). Since the string  does not wind around the $z_9$ direction, the total value of the P$(z_9)$ and F1$(z_9)$ charges is zero. Only F1$(y)$ and P$(y)$ correspond to charges measured at infinity. 
\begin{figure}[t]
	\centering
	\label{fig:LocalFig}
		\begin{tikzpicture}
			\begin{scope}[shift={(0,0)}]
				\draw[black, dashed] (-2,0) rectangle (5,5);
				\draw[blue, line width=1.2pt] (0,0) --(2.5, 5);
				%
				\begin{scope}[shift={(-0.7,2.2)}]
					\draw[blue, thick, -stealth] (0,0) -- (1,2);
					\draw[blue, thick, -stealth] (0,0) -- (1,0);
					\draw[blue, thick, -stealth] (0,0) -- (0,2);
					\draw[blue, dashed] (1,0) -- (1,2);
					\draw[blue, dashed] (0,2) -- (1,2);
					\node[align = center, centered, blue] at (1.3, 2.3) {F1};
					\node[align = center, centered, blue] at (0.5, -0.5) {F1$(y)$};
					\node[align = center, centered, blue] at (-0.1, 2.3) {F1$(z_9)$};
				\end{scope}				
				\draw[red, thick, -stealth] (1,2) -- (1+2,2-1);
				\draw[red, thick, -stealth] (1,2) -- (1+2,2);
				\draw[red, thick, -stealth] (1,2) -- (1,2-1);
				\draw[red, dashed] (1,2-1) -- (1+2,2-1);
				\draw[red, dashed] (1+2,2) -- (1+2,2-1);
				\node[align = center, centered, red] at (1+2.5,2-1.4) {$P$};
				\node[align = center, centered, red] at (1+2.6,2) {$P(y)$};
				\node[align = center, centered, red] at (1,2-1.4) {$P(z_9)$};
				\begin{scope}[shift={(-2.5,-0.5)}]
					\draw[black, thick, -stealth] (0,0) -- (2, 0);
					\draw[black, thick, -stealth] (0,0) -- ( 0,2);
					\node[align = center, centered] at (-0.5, 1) {$z_9$};
					\node[align = center, centered]at (1, -0.5) {$y$};
				\end{scope}
			\end{scope}
		\end{tikzpicture}
	\caption{Zoom in on a local piece of the fundamental string  presented in Figure~\ref{fig:F1P}.
	We decompose the string charge F1 (directed along the string direction) and momentum $P$ (directed transverse to the string) into components along the $y$ and $z_9$ directions: They source the metric and $B$-field along these directions.
	Different charge components transform into different objects upon S and T-dualization.}
\end{figure}
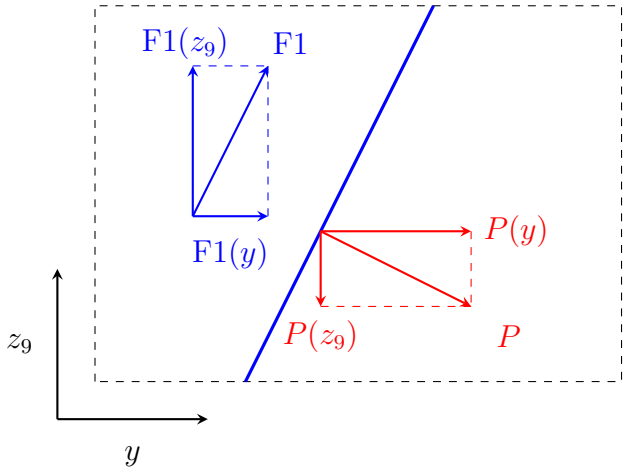

\subsubsection{NS5-P solution with local D0-D4 charges}

We now perform a series of S-dualities and T-dualities that take us to a solution with global NS5-P charges and local D0-D4 charges. 
We give here only the duality chain and the explicit expression for the final solution, leaving the  solutions obtained at intermediate steps to Appendix~\ref{app:Dualitites}.

The duality chain starts from the type-IIB solution in Equation  \eqref{eq:F1PSol}:
\begin{align}
	\label{eq:DualChain}
	&\begin{pmatrix}
		\text{F1$(y)$}\\
		P(y)\\
		\text{F1$(z_9)$}\\
		P(z_9)
	\end{pmatrix}_{\rm IIB}\hspace*{-0.4cm}
	\xleftrightarrow[]{~\text{S}~}~
	\begin{pmatrix}
		\text{D1$(y)$}\\
		P(y)\\
		\text{D1$(z_9)$}\\
		P(z_9)
	\end{pmatrix}_{\rm IIB}\hspace*{-0.4cm}
	\xleftrightarrow{~\text{T}(z_9)~}~
	\begin{pmatrix}
		\text{D2$(y, z_9)$}\\
		P(y)\\
		\text{D0}\\
		\text{F1}(z_9)
	\end{pmatrix}_{\rm IIA}\hspace*{-0.4cm}
	\xleftrightarrow{~\text{T}(z_8, z_7, z_6)~}~
	\begin{pmatrix}
		\text{D5$(y, T^4)$}\\
		P(y)\\
		\text{D3$(z_6, z_7, z_8)$}\\
		\text{F1}(z_9)
	\end{pmatrix}_{\rm IIB}\nonumber\\*
	&\qquad \xleftrightarrow{~\text{S}~}~
	\begin{pmatrix}
		\text{NS5$(y, T^4)$}\\
		P(y)\\
		\text{D3$(z_8, z_7, z_6)$}\\
		\text{D1}(z_9)
	\end{pmatrix}_{\rm IIB}\hspace*{-0.4cm}\,\,
	\xleftrightarrow{~\text{T}(z_9)~}~
	\begin{pmatrix}
		\text{NS5$(y, T^4)$}\\
		P(y)\\
		\text{D4$(T^4)$}\\
		\text{D0}
	\end{pmatrix}_{\rm IIA}\,.
\end{align}
The columns depict the objects appearing in each of the solutions, with the upper two entries denoting the charges that can be seen at infinity while the lower entries denote the local charges (which are the duals of the F1$(z_9)$ and $P(z_9)$ local charges in the solution \eqref{eq:F1PSol}). 
Above the double-headed arrows we write the duality that connects the two solutions, and show the direction along which we T-dualize. 
The subscripts of the parentheses denote the theory in which the solution exists.

At the end of the chain we obtain a solution corresponding to NS5-branes that wrap all five compact directions,  momentum $P$ along the $y$ direction, as well as D4-branes wrapping the $T^4$ and D0-branes. Note that the solution has arbitrary and equal D0 and D4 charge densities, which can either integrate to finite values or to zero. Since we are trying to construct microstate geometries for the F1-NS5-P black hole, we choose an $F(v)$ profile that does not wind along the $z_9$ direction, and which gives a solution in which the total D0 and D4 charges vanish.

Following the rules of S-dualities and T-dualities (summarized in Appendix~\ref{app:Conventions}, together with the democratic formalism \cite{Bergshoeff:2001pv} that we use to present the solution), we find that fields associated with the NS5-P solution with D0-D4 charges are given by
\begin{subequations}
	\label{eq:NS5P-D0D4}
	\begin{align}
		ds^2 &= - 2 dv\, \left[ du - \frac{\dot F(v)^2}{2}\,\left(1- \frac1{H_5}\right) \,dv\right] + H_5 \, dx^i\, dx^i+ dz^a \, dz^a\,,\\
		B_2 &= \gamma\,,\qquad e^{2\phi} = H_5\,,\\
		C_1 &= - \dot F(v)\left(1- \frac1{H_5}\right) \, dv\,, \\
		C_3 &= - \dot F(v)\, \gamma \wedge dv\,,\\
		C_5 &=- \dot F(v)\left(1- \frac1{H_5}\right) \, dv\wedge \widehat {\rm vol}_4 = C_1 \wedge \widehat {\rm vol}_4\,,\\
		C_7 & = - \dot F(v)\, \gamma \wedge dv \wedge 	\widehat {\rm vol}_4 = C_3 \wedge \widehat {\rm vol}_4\,,
	\end{align}
\end{subequations}
where  the two-form $\gamma$ is defined by
\begin{align}
	\label{eq:dgamma}
	  d\gamma \equiv *_4 d H_5\,,
\end{align}
and $ \widehat {\rm vol}_4$ denotes the volume form of the torus.
One should note that even though we started with a F1-P profile that was not isotropic along the $T^4$, through the chain of dualities \eqref{eq:DualChain} we arrive at \eqref{eq:NS5P-D0D4} where the torus only appears through its volume form.

It is useful to note that our solution exhibits the expected features.
The harmonic function $H_5$ appears in the solution in the way one expects for an NS5-brane: it multiplies the part of the metric that is transverse to the brane, it shows up in the expression for the dilaton (which diverges as one approaches the NS5-brane), and it determines the NS-NS two form which is sourced magnetically by the NS5-brane (see \eqref{eq:dgamma}).
The solution also has non-vanishing momentum, which can be read off from the $g_{vv}$ component of the metric.
This momentum arises from the non-trivial profile  function, $F(v)$, which also enters in the expression of the Ramond-Ramond gauge fields.
Since the local contribution to the momentum of the solution is proportional to  $\dot F(v)^2$,  the total momentum is always positive for any non-constant profile function.

When $F(v)$ is a constant, the solution reduces to that of a stack of NS5-branes at the origin of $\mathbb{R}^4$. When the profile function is linear in $v$, the solution describes an NS5-brane with constant D0, D4, and momentum charges. 
The D0-branes source $C_1$ electrically and $C_7$ magnetically, while the D4-branes source $C_3$ electrically and $C_5$ magnetically.
These gauge fields have the structure $C_{p+4} = C_p \wedge \widehat {\rm vol}_4$, which is a consequence of the fact that in our solution the D0 and D4 charges are locked and is related to the enhanced supersymmetry one observes when $\dot F(v)$ is constant. 

It is interesting to observe that the solution with a non-trivial $F(v)$ profile can be written in a much simpler fashion by redefining $\tilde v \equiv F(v)$. Since  $F(v)$ is periodic, and not monotonic, this re-definition is only locally well-defined, but it allows one to transform  (\ref{eq:NS5P-D0D4}) into a solution in which all the fields and metric components except $g_{u \tilde v}$ are independent of the choice of profile. Hence, the only difference between the solution with a linear $F(v)$ profile and the $v$-dependent solution with an arbitrary profile comes from multiplying  $g_{u v}$ with an arbitrary function of $v$. The fact that this multiplication transforms a solution into another solution points to the possible existence of a simple method to add null waves on certain solutions, which we plan to further explore in future work.

\subsection{Generating the F1-NS5-P solution with local D0-D4 charges}

The solution \eqref{eq:NS5P-D0D4} with a periodic $F(v)$ only has global  NS5 and P charges and can be thought of as describing a microstate of the two-charge system.
To add a third charge, we add a stack of fundamental strings on top of the NS5-P-D0-D4 solution. These strings will wrap the $S^1(y)$ circle along which the momentum is oriented, and will be  smeared along the four-torus.
To add this F1 charge we perform a duality chain on the solution in  \eqref{eq:NS5P-D0D4}, we transform it to a certain class of D1-D5-P supersymmetric solutions \cite{Giusto:2013rxa}, add an extra charge, and dualize back.

The most obvious way to dualize from the Type IIA F1-NS5-P frame to the D1-D5-P frame is to do a T-duality along the $y$ direction, followed by an S-duality. However, this supergravity duality cannot be performed on \eqref{eq:NS5F1P-D0D4}, except upon smearing the profile $F(v)$, which results in a trivial solution with no $v$ dependence. To preserve the non-trivial $v$-dependent information, one needs to T-dualize along another isometry direction.

We will use instead an isometry of the transverse space:  Rewrite the flat metric on $\mathbb{R}^4$ in the Gibbons-Hawking form \cite{Gibbons:1978tef}
\begin{align}
	\label{eq:GHForm}
	dx^i \, dx^i  &= \frac{1}{V}(d\psi + A)^2 + V\, ds_3^2\,,
\end{align}
where $\psi$ is the Gibbons-Hawking fiber, $ds_3^2$ is the line-element of flat $\mathbb{R}^3$,  $V$ is a scalar function and $A$ a one-form on this three-dimensional space, satisfying the relation \mbox{$ *_3 dA = dV$}.
Since the Gibbons-Hawking fiber is periodic, one can T-dualize along it without losing information about the local charges along the $S^1(y)$ circle, but at the cost of destroying the  asymptotic structure of the solution. However, this does not cause any problems, since we only use this duality as a tool for introducing the F1 charge: The asymptotic behavior is restored after  we dualize back to the original frame.
Hence the chain of dualities we consider is 
\begin{align}
	\label{eq:DualChain2}
	\begin{pmatrix}
		\begin{array}{c}
			\text{NS5$(y, T^4)$}\\
			P(y)\\
			\text{D4$(T^4)$}\\
			\text{D0}\\
			\hline
			\text{F1}(y)
		\end{array}
	\end{pmatrix}_{\rm IIA}\hspace*{-0.4cm}\,\,
	\xleftrightarrow[]{~\text{T}(\psi)~}~
	\begin{pmatrix}
		\begin{array}{c}
			\text{KKM$(y, T^4; \psi)$}\\
			P(y)\\
			\text{D5$(T^4, \psi)$}\\
			\text{D1$(\psi)$}\\
			\hline
			\text{F1}(y)
		\end{array}
	\end{pmatrix}_{\rm IIB}	
	\hspace*{-0.4cm}\,\,
	\xleftrightarrow[]{~\text{S}~}~
	\begin{pmatrix}
		\begin{array}{c}
			\text{KKM$(y, T^4; \psi)$}\\
			P(y)\\
			\text{NS5$(T^4, \psi)$}\\
			\text{F1$(\psi)$}\\
			\hline
			\text{D1}(y)
		\end{array}
	\end{pmatrix}_{\rm IIB},
\end{align}
where the KKM$(y, T^4; \psi)$ denotes a KKM charge with special direction $\psi$ that is distributed along the $S^1(y)$ circle and the torus. Note that the interpretations of these charges is heuristic, since the NS5-brane sits at a fixed point of the isometry of the T-duality along $
\psi$, and  the asymptotic structure is singular. 
Below the line we describe the duality chain for the fundamental string that we want to add to \eqref{eq:NS5P-D0D4}.
In the final frame (which is often called the D1-D5 frame and is commonly used in the superstrata constructions) this corresponds to adding a D1-brane wrapped along the $y$ circle. 
Since all the torus-independent supersymmetric solutions in this frame are perfectly understood \cite{Giusto:2013rxa}, we know the precise way in which to add such a D1-brane to the dual of our initial two-charge configuration, and we present the details of the calculation  in Appendix~\ref{app:Dualitites}.
 
\begin{figure}[t!]
	\centering
	\label{fig:NS5F1P-D0D4}
		\begin{tikzpicture}
			\begin{scope}[shift={(0,0)}]
				\node[align = center, centered, red] at (13.5, 0.4) {D0-D4};
				\draw[black, thick,-stealth] (1.5*3.14,-0.8)-- (2.5*3.14, -0.8);
				\node[align = center, centered] at (2*3.14, -1) {$y$};
				\draw[ForestGreen, line width=2] (0,0) -- (4*3.14,0);
				\node[align = center, centered, ForestGreen] at (13.5, 0.0) {NS5};
				\draw[red, -stealth, thick] (11,1)--(12,1);
				\node[align = center, centered, red] at (11.5, 1.2) {$P$};
				\draw[blue, line width=2] (0,-0.4) -- (4*3.14,-0.4);
				\node[align = center, centered, blue] at (13.5, -0.4 ) {F1};
			\end{scope}
			\begin{scope}[shift={(0,0.4)}]
					\coordinate (C) at (0.25pt,0);
				\foreach \x in {1,...,\N} {
					\shade [shading=axis, right color=white, left color=red, shading angle=90]
					($(\x*\T-\T,-\A)-(C)$) rectangle ++($(\T/2,2*\A)+(C)$);
					\shade [shading=axis, left color=white, right color=red, shading angle=90]
					($(\x*\T-\T/2,-\A)-(C)$) rectangle ++($(\T/2,2*\A)+(C)$);
				}
				\draw [black] plot [id=sine, domain=0:\D]
				function {\A*cos(2*pi/\T*x)};
			\end{scope}
		\end{tikzpicture}
	\caption{ A constant-time snapshot of  the periodic  $y$ direction at the origin of $\mathbb{R}^4$. 
		We have fundamental strings (F1, blue) and NS5-branes (green) wrapping the $y$ circle with momentum-carrying  D0-D4 charges densities (red density plot) living on the world-volume of the NS5-brane.
		The D0 and D4 charges have the same $y$ (or $v$) dependence, given by the profile function $F(v)$, which is necessary  for the configuration to be supersymmetric.
	}
\end{figure}
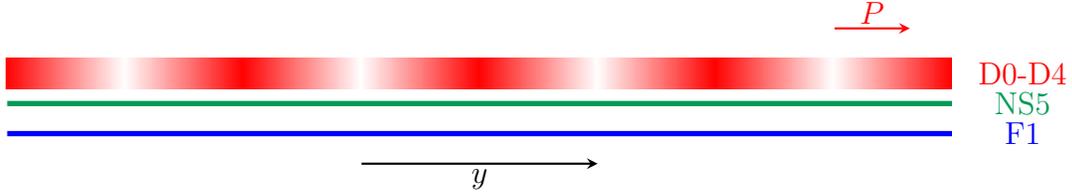

After adding the D1-brane in the D1-D5  frame \eqref{eq:DualChain2} and performing the duality transformations   backwards (from right to left), we obtain the following solution describing an F1-NS5-P system with non-trivial D0-D4 density wave, localized at the origin of  the  flat $\mathbb{R}^4$ base (see also figure~\ref{fig:NS5F1P-D0D4}):
\begin{subequations}
	\label{eq:NS5F1P-D0D4}
	\begin{align}
		ds^2 &= - \frac{2}{H_1} dv\, \left[ du - \frac{\dot F(v)^2}{2}\,\left(1- \frac1{H_5}\right) \,dv\right] + H_5 \, dx^i\, dx^i+ dz^a \, dz^a\,,\\*
		B_2 &= - \frac{1}{H_1}\, du \wedge dv+ \gamma\,, \qquad e^{2\phi} = \frac{H_5}{H_1}\,,\\*
		C_1 &=-  \dot F(v)\left(1- \frac1{H_5}\right) \, dv\,, \\*
		C_3 &=  -\dot F(v)\, \gamma \wedge dv\,,\\*
		C_5 &=- \dot F(v)\left(1- \frac1{H_5}\right) \, dv\wedge \widehat {\rm vol}_4 = C_1 \wedge \widehat {\rm vol}_4\,,\\*
		C_7 & = - \dot F(v)\, \gamma \wedge dv \wedge 	\widehat {\rm vol}_4 = C_3 \wedge \widehat {\rm vol}_4\,,
	\end{align}
\end{subequations}
where we have introduced a new harmonic function associated with the F1 charge
\begin{align}
	\label{eq:HarFun2}
	H_1(r) = 1 + \frac{Q_1}{r^2}\,,
\end{align}
and the two-form $\gamma$ is defined through \eqref{eq:dgamma}.
This solution is the main result of our construction. Note that this solution can be simplified locally in the same way as \eqref{eq:NS5P-D0D4}, by redefining the $v$ coordinate as $\tilde v = F(v)$ and seeing that all the non-trivial fluctuations along the null direction can be absorbed into a fluctuation of $g_{u\tilde v}$.

In the next section we  perform a detailed analysis of the this solution and compare it to the three-charge F1-NS5-P black-hole solution.

\section{Analysis and comparison}
\label{sec:Analysis}

In this section we compare the newly obtained  three-charge solution  \eqref{eq:NS5F1P-D0D4} to the three-charge F1-NS5-P black-hole that has a finite-size horizon. 
We begin by reviewing this black hole, focusing on the behavior of the solution near the horizon.
We then perform a similar analysis on the  solution constructed above, and compare and contrast the results. 
We find that, while the two solutions asymptotically look alike,  
they differ drastically in the near-horizon region. In the black-hole solution the singular source appearing in the harmonic function associated with the momentum is responsible for  stabilizing the $y$-circle thus giving rise to an event  horizon with a finite area.
This does not happen in the new solution \eqref{eq:NS5F1P-D0D4}, where the momentum is produced by the fluctuations of the local D0 and D4 charges, whose corresponding function remains finite at the location of the F1 and NS5-brane sources. As a consequence, the $y$-circle pinches off and the horizon area vanishes. The existence of our solution indicates that if one considers all the degrees of freedom of the system, an event horizon does not form even when the system has no transverse fluctuations.

\subsection{The F1-NS5-P three-charge black hole}

The F1-NS5-P three-charge black hole 
is obtained by superimposing a stack of NS5-branes (wrapping $S^1(y)\times T^4$) and a stack of F1-strings  (wrapping $S^1(y)$), both of which are located at the origin of $\mathbb{R}^4$, and allowing for additional momentum charge in the $y$ direction \cite{Tseytlin:1996bh}. This yields the solution:%
\footnote{Throughout section~\ref{sec:Analysis} we are working with string-frame metrics, unless explicitly stated otherwise.}
\begin{subequations}
	\label{eq:NS5F1PBlackHole}
	\begin{align}
		ds^2 &= - \frac{2}{H_1}\, dv\,\left(du + \frac{\cF}{2}\, dv\right) +  H_5 \, dx^i\, dx^i+ dz^a \, dz^a\,,\\*
		B_2 &= - \frac{1}{H_1}\, du\wedge dv + \gamma\,, \qquad 	e^{2\phi} = \frac{H_5}{H_1}\,,
	\end{align}
\end{subequations}
with all other fields vanishing. 

The harmonic functions associated to the NS5-branes and F1-strings, $H_5$ and $H_1$, are given by the expressions \eqref{eq:HarFun} and  \eqref{eq:HarFun2}.
Furthermore, the magnetic component of $B_2$, which is sourced by the NS5-branes is given by the expression \eqref{eq:dgamma}.
The harmonic function associated to the momentum, $\cF$, has a $\delta$-function source at the origin of $\mathbb{R}^4$, whose strength is proportional to the momentum charge as measured at spatial infinity, $Q_P$:
\begin{align}\label{eq:MomHarm}
 \cF = - \frac{2 \,Q_P}{r^2}\,.
\end{align}

In the backreacted solution, there is an event horizon at $r = 0$. To calculate its area one needs to look at the size of the orthogonal dimensions as one approaches it. 
One can show  that the radius of the $S^1(y)$ circle at an arbitrary value of $r$ is
\begin{align}
	\label{eq:BHyRad}
	R_y(r) = \sqrt{\frac{Q_P + r^2}{Q_1 + r^2}}\, R_y\,,
\end{align}
where, as before, $R_y$ denotes the value of this radius at infinity. We can see that the $y$-circle remains finite in size as we approach the horizon at $r  = 0$.
Combining this with the finite size of the $S^3$ of the $\mathbb{R}^4$, we find that \eqref{eq:NS5F1PBlackHole} has a non-zero horizon area.
This is a direct consequence of the stabilization of the $S^1(y)$ circle at the location of the horizon, caused by  the balancing between the effect of the momentum, which exerts a centrifugal force towards a large radius, and the tension of the branes wrapping the circle, which try to  shrink it.
In the absence of momentum ($Q_P=0$), one can see from \eqref{eq:BHyRad} that the $S^1(y)$ circle wrapped by the NS5-branes and F1-strings pinches off as $r \to 0$ and thus the horizon area vanishes.

Finally, we note that the metric is actually smooth at the horizon, and it can be smoothly continued across it.  As one would expect, the curvature  invariants remain finite: 
\begin{subequations}
	\label{eq:CurvInv}
	\begin{align}
		R& = - 20 \frac{Q_1 - Q_5}{Q_1\, Q_5^2}\, r^2 + \cO\left(r^3\right)\,,\\
		R_{\mu\nu}\,R^{\mu\nu} &=  \frac{24}{ Q_5^2} + \cO\left(r^2\right)\,,\\
		R_{\mu\nu\rho\sigma}\,R^{\mu\nu\rho\sigma}&=  \frac{24}{ Q_5^2} + \cO\left(r^2\right)\,.
	\end{align}
\end{subequations}

\subsection{The new three-charge solution with local D0-D4 charges}

We can write the metric of our new solution \eqref{eq:NS5F1P-D0D4} as
\begin{subequations}
	\label{eq:NS5F1P-D0D4Metric}
	\begin{align}
		ds^2 &= - \frac{2}{H_1} dv\, \left[ du - \frac{\dot F(v)^2}{2}\,\left(1- \frac1{H_5}\right) \,dv\right] + H_5 \, dx^i\, dx^i+ dz^a \, dz^a\,,\\*
		&= \frac{1}{H_1}\left[-dt^2 +  dy^2+ \frac{\dot F(v)^2}{2}\,\left(1- \frac{1}{H_5}\right)\, \left(dt+ dy\right)^2\right]+  H_5 \, dx^i\, dx^i+ dz^a \, dz^a\,,
	\end{align}
\end{subequations}
where we used  \eqref{eq:NullCC} to obtain the second line. 
If the harmonic functions $H_1$ and $H_5$ contain a constant, the geometry is asymptotically flat $\mathbb{R}^{4,1} \times S_y  \times T^4$. 
The main difference with the black hole comes from the behavior of the $g_{vv}$ component of the metric,  which contains the information about the momentum of the system. 
In contrast to \eqref{eq:NS5F1PBlackHole}, this metric does not contain a freely choosable harmonic function, $\mathcal{F}$, with an independent charge $Q_P$. Rather, the momentum is encoded in the profile $F(v)$ and the combination $(1- H_5^{-1})$, which, as already mentioned, is finite everywhere in the base space.
This is because the momentum is carried in a fundamentally different way compared to the black-hole solution. The finiteness of $(1- H_5^{-1})$ suggests an absence of a localized source for the momentum. This is in conflict with the naive ``NS5 world-volume intuition," according to which the momentum is sourced by longitudinal fluctuations of the D0 and D4 densities inside the NS5-brane world-volume, and hence it should also be sourced at the location of the NS5-brane. Of course, the NS5 world-volume intuition ignores back-reaction, so it is not the appropriate intuition for the the full supergravity solution. But it is rather puzzling that other aspects of this world-volume intuition are described correctly in supergravity, while this particular aspect is not.

\subsubsection{The asymptotics}

Despite the absence of a singular source, one can calculate the value of the momentum along the $y$ direction in this solution from the asymptotic expansion \cite{Myers:1986un, Heidmann:2019xrd}:
\begin{align}
	g_{vv} \approx \frac{1}{r^2}\left( 2Q_P+ \text{\textit{oscillating terms}}\right) + \cO\left(r^{-3}\right)\,.
\end{align}
Thus we can read off
\begin{align}
	\label{eq:gvv_asymptotics}
	g_{vv} = \frac{\dot F(v)^2}{H_1}\,\left(1- \frac1{H_5}\right) \approx \frac{Q_5 \, \dot F(v)^2}{r^2} +  \cO\left(r^{-3}\right)\,,
\end{align}	
from which we extract the non-oscillating part by averaging over the $y$-circle: 
\begin{align}
	\label{eq:MomSol}
	Q_P = \frac{Q_5}{2}\, \frac{1}{\sqrt2 \pi R_y}\,\int_0^{\sqrt2\pi R_y}\, \dot F(v)^2\, dv\,.
\end{align}	
Note that if the profile function admits a decomposition as a Fourier sum 
\begin{align}
	\label{eq:FvExp}
	F(v) = R_y \,a_0+  R_y\sum_{n=1}^{\infty}\,\left[\frac{a_n}{n}\, \cos\left( \frac{\sqrt 2 n v}{R_y}\right) + \frac{b_n}{n}\, \sin\left( \frac{\sqrt 2 n v}{R_y}\right)\right]\,,
\end{align}
then one can evaluate the integral in \eqref{eq:MomSol} and obtain
\begin{align}
	Q_P = \frac{Q_5}{2}\,\sum_{n=1}^\infty\left(a_n^2 + b_n^2\right)\,.
\end{align}	
Thus, different solutions in the family we constructed (\ref{eq:NS5F1P-D0D4}), parameterized by different profile functions $F(v)$, have the same  asymptotic momentum charge, $Q_P$, as the black hole (\ref{eq:MomSol}).
However, while the $g_{vv}$ component of the black-hole solution only contains a harmonic function proportional to $Q_P$
\begin{align}
	g_{vv}^{\mathrm{BH}}= \frac{1}{H_1} \frac{2 \,Q_P}{r^2} \,,
\end{align}
the metric of our solutions deviate from that of the black hole at higher order in the asymptotic expansion in $r$, because of the $(1- H_5^{-1})$ term in $g_{vv}$ (\ref{eq:gvv_asymptotics}):
\begin{align}
\label{eq:gvv_expansion}
	g_{vv}(v) = \frac{\dot{F}(v)^2}{H_1} \( \frac{Q_5}{r^2} - \frac{Q_5^2}{r^4} + \mathcal{O}\(r^{-6}\) \) \,.
\end{align}
Averaging (\ref{eq:gvv_expansion}) over $v$ suggests that the higher multipoles of our solutions may be different from those of the black hole: 
\begin{align}
	\langle g_{vv}\rangle_v \equiv \frac{1}{\sqrt{2}\pi R_y}\int_0^{\sqrt{2}\pi R_y}g_{vv}(v)\,dv 
	= \frac{1}{H_1} \( \frac{2 \,Q_P}{r^2} - \frac{2 \, Q_5 \, Q_P}{r^4} + \mathcal{O}\(r^{-6}\) \) \,.
\end{align}
Hence, our solution deviates from the black-hole metric via $\frac{Q_5Q_P}{r^4}$ and higher terms in $g_{vv}$, which  indicates that the momentum wave of the microstructure in the backreacted solution develops a finite size. This will be further confirmed in Section \ref{sec:nh-limit}.

\subsubsection{The vanishing-area horizon}

Much like in the two-charge F1-NS5 solution, one finds that $g_{tt}$ goes to zero at $r = 0$, the location of the pole of the brane harmonic functions. Furthermore, the curvature invariants are finite at this point and are equal to those of the F1-NS5 two-charge solution\footnote{One should remember that the near-brane limit of the two-charge solution is, locally, like Poincar\'e AdS$_3$ $\times  S^3$, and so the curvature invariants are all well-behaved.  What makes the solution singular is the fact that the $S^1$ pinches off in the $r \to 0$ limit, where  $g_{tt}$ also vanishes.} and those of the F1-NS5-P three-charge black hole \eqref{eq:CurvInv}.
The crucial difference comes from behavior of the  length of the $y$-circle near the brane sources, which we calculate using  \eqref{eq:NS5F1P-D0D4Metric} 
\begin{align}
	\label{eq:LyCircle}
	L_y = \sqrt{\frac{2}{H_1}}\int_0^{\sqrt 2\pi R_y}\sqrt{1+ \frac{\dot F(v)^2}{2}\left(1- \frac1{H_5}\right)}\,dv\approx r\,\sqrt{\frac{2}{{Q_1}}}\int_0^{\sqrt2\pi R_y}\sqrt{1+ \frac{\dot F(v)^2}{2}}\,dv\,,
\end{align}
where we have expanded around $r = 0$. 
Since the integrand is a strictly positive function, we find that near the origin the $y$-circle pinches off, despite the fact that the solution has a non-trivial momentum along that direction. 
One can show that,  as $r\to 0$, all other dimensions are finite in size.%
\footnote{One can show that the three-sphere which appears in the base space has an area of ${\rm Area}(S^3) = 2 \pi^2 \,\left(r^2\,H_5\right)^{\frac32} \approx 2\pi^2 Q_5^{\frac32}$, where we have expanded near $r = 0$. 
	Furthermore,  the volume of the $T^4$ is independent of $r$ and is taken to be finite. Then the string-frame area of the would-be horizon is $A_H = L_y \, {\rm Area}(S^3) \, {\rm Vol}(T^4)$, which vanishes as one approaches the brane sources because of the pinching of the $y$-circle.}
Therefore,  \eqref{eq:NS5F1P-D0D4} has a singularity that can be thought of as a zero-area horizon. This is the same type of singularity as in the F1-NS5 or D1-D5 two-charge solutions.
Our new solution is thus very peculiar: For a non-trivial profile $F(v)$, we can see from \eqref{eq:MomSol} that it contains momentum along with F1 and NS5 charges, making it a three-charge solution. On the other hand, one can see from \eqref{eq:LyCircle} that the $y$-circle shrinks at the origin, which gives rise to a singularity of the type present in two-charge solutions.

\subsubsection{The near-horizon behavior - a first pass}
\label{sec:nh-limit}

There exist two ways to analyze the near-horizon behavior of the solution. One can, as we discuss in this subsection,  focus on the region where
\begin{align}
	\label{eq:SmallrExp}
	r^2 \ll Q_1, Q_5\,.
\end{align}
By expanding \eqref{eq:NS5F1P-D0D4} in small $r$, one can probe the solution in the vicinity of the brane sources. 
The expansion of the metric is, up to order $\cO(r^2)$, given by:
\begin{align}
	\label{eq:near-horizon_IIA}
	ds^2 \,=\, \sqrt{\frac{Q_5}{Q_1}}\, \left[ - \frac{2\, r^2}{\sqrt{Q_1 \, Q_5}}\, dv \left( du - \frac{\dot F^2(v)}{2} dv\right) + \frac{\sqrt{Q_1 \, Q_5}}{r^2}\, dr^2 + \sqrt{Q_1 \, Q_5}\, d\Omega_3^2\right]+d\hat s_4^2\,,
\end{align}
which is locally AdS$_3\times S^3\times T^4$, as can be seen more explicitly by introducing a new coordinate 
\begin{align}
	\label{eq:LocalTrans}
	w \equiv  u - \int\frac{\dot F(v)^2}{2} \, dv\,, \qquad 
	dw = du - \frac{\dot F(v)^2}{2} \, dv\,.
\end{align}
Thus, near the brane sources, the solution is locally simply empty AdS.
The transformation \eqref{eq:LocalTrans} removes the metric component $g_{vv}\propto \dot F^2(v) \, r^2$, which is the only term  in the near-horizon region sensitive 
to $\dot F^2(v)$.
This metric component vanishes at $r \to 0$, but grows as $r^2$ with increasing radius.
Therefore, it does not vanish at the boundary of AdS$_3$ ($r \to \infty$), but corresponds to a non-trivial deformation of the boundary metric.

\begin{figure}[t!]
	\includegraphics[width=\linewidth]{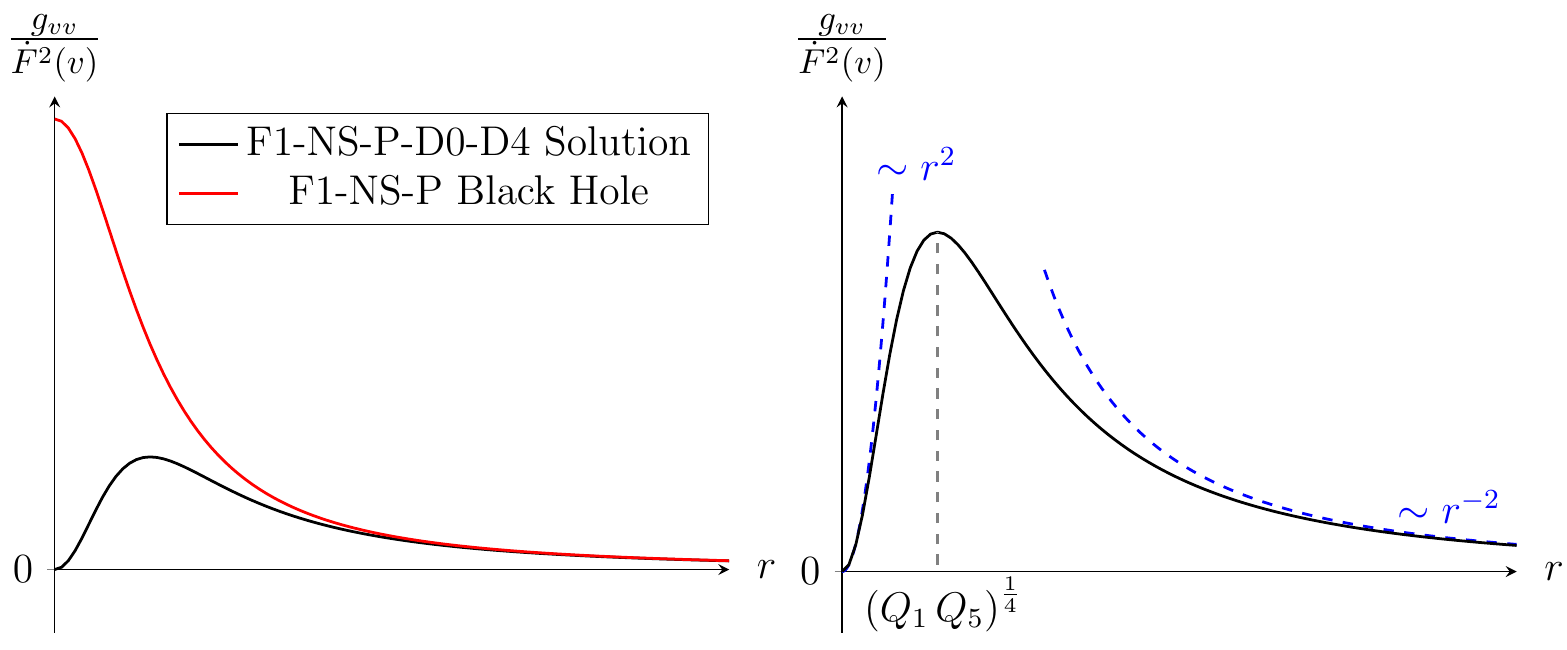}
	\caption{The schematic behavior of the metric component $g_{vv}$ as a function of the radial coordinate. 
		On the left is the plot of the exact expression (modulo the $\dot F^2(v)$ function)  for the new F1-NS5-P-D0-D4 solution \eqref{eq:gvv_asymptotics} (black) and the  F1-NS5-P black hole (red). 
		The momentum charge of the latter is taken to be such that the asymptotic behavior of the two solutions match.
		In the bulk the two solutions differ significantly: At $r =0$ the black hole has a finite value for $g_{vv}$ which is related to the finite size of the horizon, while in the new solution this metric component vanishes and the $S^1(y)$  circle pinches off. 
		On the right, we have a close-up of the  solution with local D0-D4 charges,  superposed  with the asymptotic and near-brane behavior in blue. 
		The momentum is localized away from the brane sources, with the maximum  at $r^2 = \sqrt{Q_1 \, Q_5}$. 
	}
	\label{fig:Plotgvv}
\end{figure}

The growing behavior of $g_{vv}$ as one is increasing the radius implies that the momentum is not localized in the interior of the AdS region.
Since the asymptotically-flat solution \eqref{eq:NS5F1P-D0D4} contains non-vanishing momentum charge,  the momentum wave must be located in the  transition zone between the AdS$_3$ near-horizon region and the  flat space region. 
This explains why our new solution has a momentum that can be measured at infinity \eqref{eq:MomSol}, despite the absence of a no momentum-charge source at $r=0$.
Indeed, as can be seen from figure~\ref{fig:Plotgvv}, which depicts the $g_{vv}$ for arbitrary values of $r$, \eqref{eq:near-horizon_IIA} captures only the leading near-horizon behavior but fails to capture the asymptotic fall-off. 
Furthermore, in the string frame the maximum of  $g_{vv}$ is located at $r^2 = \sqrt{Q_1 \, Q_5}$, providing further evidence that the momentum wave is localized in the transition region between AdS$_3$ and flat space.

Finally, let us note that the metric \eqref{eq:near-horizon_IIA} does not correspond to the results from the heuristic method of taking a near-horizon limit by ``dropping the 1'' in the harmonic functions. 
This method gives a metric which has an additional term:
\begin{align}
	\label{eq:DecLimDrop1}
	ds^2 \,=\, \sqrt{\frac{Q_5}{Q_1}}\, \left[ - \frac{2\, r^2}{\sqrt{Q_1 \, Q_5}}\, dv \left( dw +  \frac{\dot F^2(v)\,r^2}{2\, Q_5} dv\right) + \frac{\sqrt{Q_1 \, Q_5}}{r^2}\, dr^2 + \sqrt{Q_1 \, Q_5}\, d\Omega_3^2\right]+d\hat s_4^2\,,
\end{align}
where we have used the shifted coordinate \eqref{eq:LocalTrans}. This metric corresponds holographically to a deformation of AdS$_3\times S^3\times T^4$ with a non-normalizable mode corresponding to an irrelevant operator of the dual CFT. Furthermore, the metric is no longer locally AdS: the additional term in $g_{vv}$ that scales as $r^4$ and diverges at the boundary of AdS cannot be reabsorbed by a coordinate transformation. 

This deformation of the metric is accompanied by a non-vanishing deformation of the RR gauge fields: 
\begin{align}
	\label{eq:NHlimit2}
	C_1 = \left(1- \frac{r^2}{Q_5}\right)\, \dot F(v)\, dv\,,\qquad C_3 = -Q_5\, \dot F(v)\, \gamma'\wedge dv\,,
\end{align}
and all higher order forms can be obtained by using the self-duality conditions \eqref{eq:DemFormSD}.
In $C_3$ we have used the fact that when writing $\IR^4$ in spherical coordinates, $d\gamma = *_4 dH_5 = 2 Q_5\, {\rm vol}\left(S^3\right)$. 
Thus it is convenient to define a new, ``bare'', two-form $\gamma'$ such that $d\gamma' \equiv 2\, {\rm vol}\left(S^3\right)$.
It then naturally follows that $C_3$ remains unchanged in the near-horizon expansion, since it is independent of the radial coordinate.
Finally, the NS-NS gauge field is the same as in the standard decoupling limit and the corresponding field strength supports the AdS$_3\times S^3$ structure. 

\subsubsection{The near-horizon behavior - a  second pass}
\label{sec:nh-limit2}

Another way of decoupling the near-horizon region from the asymptotically flat region and obtain a background that is holographically dual to the low-energy physics of a brane system is to take a double-scaling limit \cite{Maldacena:1997re} involving $\alpha'$ and the transverse radial direction.
To do this we need to first express the charges appearing in the supergravity solution, $Q_1$ and $Q_5$, in terms of the moduli and the quantized numbers of F1 strings, $N_1$, and NS5-branes, $N_5$: 
\begin{align}
	\label{eq:SugraCharges}
	Q_1 = \frac{g_s^2\, \alpha'^3}{V_4}\, N_1\,, \qquad Q_5 = \alpha'\, N_5\,,
\end{align}
where $g_s$ is the string coupling constant, $\alpha'$ is the Regge slope, and $V_4$ is the coordinate volume of the four-torus divided by $(2\pi)^4$. 
The double scaling limit is \cite{Maldacena:1997re}
\begin{align}
	\label{eq:DecouplingLimit}
	\alpha' \to 0\,, \qquad U \equiv  \frac{r}{\alpha'}= \text{fixed}\,, \qquad v_4 \equiv \frac{V_4}{\alpha'^2} =\text{fixed}\,, \qquad g_6 \equiv  \frac{g_s}{\sqrt{v_4}} = \text{fixed}\,,
\end{align}
and it yields the ten-dimensional string frame metric: 
\begin{align}
	\label{eq:MetricDecoupling}
	\frac{ds^2}{\alpha'} = N_5 \Biggr[ - \frac{2\, U^2}{g_6^2 \, N_1\, N_5}\, dv \left( du - \frac{\dot F^2(v)}{2} dv\right) + \frac{dU^2}{U^2}\, dU^2  +  d\Omega_3^2\Biggr]+ dz^a \, dz^a\,.
\end{align}
This result is consistent with the near-brane expansion of the metric \eqref{eq:near-horizon_IIA}, provided one makes the substitutions $Q_1\to g_6^2\, N_1$ and $Q_5 \to N_5$.
Thus, as before, the metric in the decoupling limit corresponds to locally empty AdS, with a deformation that is non-trivial at the asymptotic boundary.
Performing the same scaling on the gauge fields in the solution \eqref{eq:NS5F1P-D0D4}, one finds that the NS-NS two-form becomes such that the corresponding field strength is comprised of a part proportional to the volume form of AdS$_3$ and a part proportional to the volume form of $S^3$.
On the other hand, the RR gauge fields $C_p$ are such that all field strengths, $F_{p+1}$, vanish in this limit. 

It is important to note that the double scaling limit \eqref{eq:DecouplingLimit} and the near-brane expansion considered in \eqref{eq:near-horizon_IIA} lose all information about the harmonic function $H_5$ appearing in $g_{vv}$ and about the nontrivial RR fields of the solution. It is interesting to try to construct a decoupling limit which does not erase this information. It is not hard to see that such a limit combines \eqref{eq:DecouplingLimit} with a scaling of the null coordinates defined in \eqref{eq:LocalTrans},  
while keeping fixed 
\begin{align}
	\label{eq:AddScaling}
	d\tilde v \equiv \sqrt{\alpha'} dv = \text{fixed}\,, \qquad d\tilde w \equiv \frac{dw}{\sqrt{\alpha'}} = \text{fixed}\,.
\end{align}
This results in a metric%
\footnote{Note that despite the scaling \eqref{eq:AddScaling} we keep $\dot F(v)$ fixed.
This can be achieved by scaling $F(v)$ in  a way which  cancels out the scaling of $v$ coming from the differentiation.} 
\begin{align}
	\label{eq:NewDecLim}
	\frac{ds^2}{\alpha'}  = N_5 \left[ - \frac{2\, U^2}{g_6^2 \, N_1\, N_5}\, d\tilde v \left( d\tilde w +  \frac{\dot F^2(\tilde v)\,U^2}{2\, N_5} d\tilde v\right) + \frac{dU^2}{U^2} + \, d\Omega_3^2\right]+ dz^a \, dz^a\,,
\end{align}
corresponding to a non-trivial deformation of AdS$_3\times S^3\times T^4$.
We also find the non-trivial RR gauge fields
\begin{align}
	\label{eq:NewDecLimGauge}
	C_1 = - \frac{U^2}{N_5}\, \dot F(\tilde v)\, d\tilde v\,, \qquad C_3= - N_5\, \dot F(\tilde v)\, \gamma' \wedge d\tilde v\,,
\end{align}
where in writing the latter expression we again used the two-form $\gamma'$, as  defined in \eqref{eq:NHlimit2}.
All higher-order forms can be obtained from these by using the democratic formalism.
It is interesting to observe that despite the non-trivial scaling of the coordinates $\tilde w$ and $\tilde v$, the final result matches the one obtained by simply ``dropping the 1'' in the harmonic functions \eqref{eq:DecLimDrop1}, if one appropriately identifies coordinates and moduli of the two solutions.

Finally, let us note that the same results can be obtained by another scaling limit which is more commonly used in the F1-NS5-P system \cite{Martinec:2017ztd, Martinec:2018nco, Bufalini:2021ndn}.
Begin by defining dimensionless coordinates $\tilde u \equiv {u}/{R_y}$ and $\tilde v \equiv {v}/{R_y}$.
Then one takes the  AdS$_3$ decoupling limit%
\footnote{For the F1-NS5-P system there exists an additional linear-dilaton region \cite{Kutasov:2001uf} which is obtained by taking only $g_s \to 0$ while keeping the ratio $r/g_s$ fixed. As can be seen from \eqref{eq:SugraCharges}, this limit focuses on the region of spacetime where $Q_1 \ll r^2 \ll Q_5$. 
We are interested in the scaling which accesses the region \eqref{eq:SmallrExp}, which is achieved by the scaling described in the main text.
We would like to thank David Turton and Soumangsu Chakraborty for helpful discussions on this point. } 
by scaling $g_s \to 0$ and $R_y \to \infty$, while keeping fixed the supergravity charges, $Q_1$ and $Q_5$, the coordinates $\tilde u$, $\tilde v$, and $r/g_s$,  and the remaining string moduli. 
In practice, we can implement this limit by making the  replacements \cite{Martinec:2018nco}
\begin{align}
	\label{eq:DecoupRescaling}
r \to \epsilon\, r\,, \qquad R_y \to \frac{R_y}{\epsilon}\,,
\end{align}
followed by sending $\epsilon \to 0$.%
\footnote{Again we keep $\dot F(v)$ fixed in this scaling.}
One finds that the resulting metric is exactly equal to  \eqref{eq:near-horizon_IIA}, obtained by the near-brane expansion of the full asymptotically flat geometry. 
If, on the other hand, one first performs the transformation \eqref{eq:LocalTrans}, defines $\tilde w \equiv {w}/{R_y}$, and, in addition to \eqref{eq:DecoupRescaling}, scales
\begin{align}
	\tilde w \to \epsilon\, \tilde w\,, \qquad \tilde v \to \frac{\tilde v}{\epsilon}\,, 
\end{align}
then the $\epsilon \to 0$ limit  yields the solution obtained by ``dropping the 1'' in the Harmonic functions \eqref{eq:NewDecLim}.

\subsection{Supersymmetries and singularities}

Since our  NS5-P-D0-D4 solution is a dual of the  F1-P string, it must have eight supersymmetries, which are identical to the common supersymmetries preserved by NS5-branes and a momentum wave.  Moreover,   if one zooms in locally, the function,  $F(v)$, becomes approximately  linear in $v$, and the resulting solution has 16 supersymmetries.
One can also confirm this by directly calculating the brane projectors, like in \cite{Bena:2011uw}.
Alternatively, this can be seen by noting that such a linear solution comes from dualizing a tilted fundamental string boosted orthogonally,  or equivalently, by uplifting to 11 dimensions, where the linear system becomes an M5-brane with orthogonal momentum, as depicted in  Figure~\ref{fig:MTheoryPic}. 
Both such configurations preserve 16 supersymmetries.

It is natural to ask how the NS5-P-D0-D4 solution can preserve the same supersymmetries as the NS5-P system, despite the presence of D0 and D4 densities. This is achieved because the D0 and D4 densities have the same distribution on the $S^1(y)$-circle, which makes their joint contribution to the supersymmetry projector  compatible with the Killing spinors preserved by NS5-branes and momentum. This phenomenon was observed in the construction of the magnetube \cite{Bena:2013ora}, and it is not hard to see that if one T-dualizes our solution twice along the D4-brane world-volume, one obtains an NS5-D2-D2-P brane configuration that uplifts to the M5-M2-M2-P magnetube of \cite{Bena:2013ora}.

Upon adding F1-strings to the NS5-P-D0-D4 solution, the supersymmetry is reduced to half. Thus, the resulting solution has globally four supercharges, but if one zooms near the source (or considers a solution with a linear $F(v)$)  the number of supercharges is enhanced to eight. This is consistent with the fact that the singularity in this solution is the same as that of a two-charge single-center solution.

\section{Conclusion and discussion}
\label{sec:Discussion}


The Fuzzball and Microstate Geometry Programmes exist precisely because string theory and supergravity have a rich variety of degrees of freedom that can be used to evade the formation of horizons. 
A recent, but illustrative example is the long-term trapping \cite{Eperon:2016cdd} near evanescent ergosurfaces which was believed to lead to Aichelburg-Sexl shockwaves and horizon formation.
However, a more detailed analysis showed that this would actually result in scrambling into more and more typical modes of the solution \cite{Marolf:2016nwu}. 
Furthermore,  the extremely long-term trapping needed to create singularities requires sub-stringy wavelengths for the modes  \cite{Bena:2020yii}.   In short, the stringy degrees of freedom are activated before horizons develop and one must explore the full range of supergravity and stringy phase space or one risks mimicking the limitations of General Relativity and concluding that  horizons are inevitable.  

In this work we  examined another manifestation of this phenomenon: In the D1-D5 frame, a family of smooth, three-charge Microstate Geometries (the \textit{superstrata} family) appears to develop a horizon in  the limit of vanishing angular momentum ($a\to0$).  We have now given strong evidence that \textit{the horizon only emerges because one has neglected degrees of freedom that are essential in the $a\to0$ limit}.
Indeed,  we incorporated some of these degrees of freedom by introducing D0- and D4-brane densities  in the Type IIA F1-NS5 frame and showed that these resulted in a solution that has a vanishing horizon area.

We have also understood that reason behind the failure of the na\"ive intuition according to which $a\to0$ D1-D5-P superstrata appear to collapse into a black hole. The momentum of these superstrata is only carried by D1 and D5 dipole-charge distributions \cite{Niehoff:2013kia,Bena:2015bea} that are compressed to zero size in  the $a\to0$ limit.%
\footnote{Furthermore, in bubbling solutions\cite{Bena:2005va, Berglund:2005vb} the momentum charge comes from the non-trivial dipole fluxes, which also vanish when $a\to 0$.}  If one takes into account all possible momentum carriers, no such collapse happens. 

Indeed, the D1-D5 configuration on which one builds the microstate geometries comes from dualizing an F1-string with momentum, and since the F1-string only carries momentum waves that are transversely polarized \cite{Lunin:2001fv}, this configuration has finite size. By contrast, we find that NS5-branes can carry momentum also through longitudinal fluctuations, via a non-trivial profile of world-volume fluxes corresponding to D0- and D4-brane densities. It is this fact that  allows us to construct 3-charge zero-horizon-area solutions, despite the NS5-branes being localized at a single point in the $\IR^4$ base space. Hence, our solutions are $SO(4)$ singlets under rotations on the $\mathbb{R}^4$, exactly as the usual three-charge black hole solution.

An interesting observation, which only emerges from analyzing the full supergravity solution, is that the momentum ``carried'' by the D0 and D4 charge densities inside the NS5 world-volume is {\em not} localized near the NS5-brane source, but resides in the transition region between the near-horizon AdS$_3\times S^3$ and the asymptotically flat region. As such, this momentum cannot prevent the $S^1(y)$ wrapped by the F1-strings and the NS5-branes from collapsing at the location of the brane sources, which in turn causes the horizon area to vanish. %

As we remarked earlier, there is an important distinction between microstate solutions and degenerate microstate solutions.  Both have vanishing horizon area, but the former represent  pure states, whereas the latter  encode a  large number of microstates.  The singularities of two-charge solutions, like the F1-NS5 singularity, or the D1-D5 singularity, and the singular core of our F1-NS5-P-D0-D4 solution are, in this sense, degenerate microstate solutions, and their cores represent  ensembles of microstates that have neither the charges nor the degrees of freedom to create a macroscopic horizon. 

Degenerate microstate solutions are also required to have microstructure that can be understood using string theory.   Resolving the microstructure of the singular  D1-D5 system was the focus of the original fuzzball program \cite{Lunin:2001fv,Lunin:2002iz}.  More recently, our understanding of the microstructure of the F1-NS5 system has been greatly advanced using world-sheet methods \cite{Martinec:2017ztd,Martinec:2018nco,Martinec:2019wzw,Martinec:2021vpk, Bufalini:2021ndn}. 

Our work has enriched the ``landscape'' of superstrata  by expanding the range of momentum carriers on the branes.  As we have seen, the addition of the D0-D4 excitations reveals how the fuzzball paradigm works even in the singular corners \cite{deBoer:2008fk,Li:2021gbg,Li:2021utg} of the moduli space.   This also suggests several interesting areas for further investigation: we expect that there are whole new classes of microstate geometries that come from the geometric transition of our degenerate microstate solutions.  Another intriguing question is whether there are such transitions that only involve  the $T^4$, and achieve this in a way that preserves the space-time $SO(4)$ invariance and the vanishing angular momentum. 

It would also be interesting to see, in detail, how the solutions obtained in this paper  emerge as a limit of smooth microstate geometries. In particular, one should be able to construct superstrata, with $a >0$, that contain both ``standard'' momentum carriers and  D0-D4 momentum carriers. In such a generalized superstratum with $a >0$,  the $y$-circle should pinch off smoothly, making a smooth cap at the bottom of a long BTZ-like throat. It would be interesting to construct this Type-IIA superstratum with F1-NS5-P charges, and to explore its  $a \to 0$ limit and the relation of this limit to the solutions we construct in this paper.

In particular, if there exist Type IIA superstrata that limit to our solutions, there is then the question of what happens to the long BTZ throat.  Do our solutions emerge in the center of a cap at the bottom of a long throat, or does the throat become much shallower?    Indeed, this is directly related to the results presented in Section \ref{sec:nh-limit}, where we showed that in the full supergravity solution, the momentum charge comes from modes localized in the junction  between the near-horizon AdS$_3\times S^3$ region and the asymptotic flat space. In a generalized superstratum, with D0-D4 momentum carriers and with $a>0$, we would still expect that, like in the original superstrata, all the momentum waves should localize in a band that creates the transition between the horizonless cap and the long \mbox{AdS$_2\times S^1$} region of the BTZ throat.   It would be very interesting to see whether and how the location of the momentum waves shifts in the $a \to 0$ limit of the generalized superstratum.

Even though our solutions have the same spherical symmetry as a single-center black hole with the same charges, their asymptotic expansions are different. This happens because the momentum is carried by null waves located at the top of the AdS$_3\times S^3$ throat, and hence there is no limit of our solutions where they approach those of the black-hole solution to arbitrary precision.  
This makes them different from the usual microstate geometries which have a ``scaling'' parameter controlling the depth of the throat, that can be tuned so that their metric and the gravitational multipoles approach those of the black hole \cite{Bena:2020uup,Bah:2021jno}. Our new solutions do not have such a parameter and hence we expect them to have a metric whose asymptotics differs from that of the black-hole solution at higher orders in the radial distance. 
Furthermore, although the extra fields in our solutions  fluctuate along a null coordinate, they all contribute to the  metric with the same sign. Hence, even if one considers an ensemble of our new solutions with D0-D4 modes, these features will not average to zero, and the $1/r$-expansion will still differ from that of the black hole.

The location of the momentum also presents a puzzle in terms of the dual CFT picture. 
As discussed in the introduction, we expect that, in the $a \to 0$ limit, the state dual to the superstratum consists of momentum-carrying $(0,0)$ strands and no $(+,+)$ strands. 
However, in our solution taking the standard decoupling limit results in a locally AdS$_3\times S^3\times T^4$ spacetime, \eqref{eq:MetricDecoupling} with a deformation to the metric at the boundary of the spacetime. 
Furthermore, performing an alternative scaling, one can obtain an AdS$_3\times S^3\times T^4$ solution deformed with an non-normalizable momentum-carrying mode dual to an irrelevant deformation of the CFT.
If, as mentioned above, in a generalized superstratum one were to find some microstructure at the center of a smooth cap, then there should exists an equivalent description in the dual CFT. 
Establishing the precise holographic  dictionary for both the new microstate solution and potential generalized superstrata, is thus of great interest.  

From a technical point of view, constructing   generalized superstrata requires solving a new set of non-trivial BPS equations. 
From the perspective of six-dimensional supergravity, the ten-dimensional fields sourced by the D0 and D4 charge densities are encoded in a U(1) gauge field. Furthermore, the equations governing six-dimensional supersymmetric solutions with tensor and vector gauge fields were derived in \cite{Cariglia:2004kk}. 
It is important to remember that the construction of the original superstrata relied on the hidden linear structure of the BPS equations of six-dimensional supergravity with tensor fields, but no gauge fields \cite{Bena:2011dd,Bena:2015bea}.
In an upcoming paper \cite{linear-structure} we will show that such a linear structure persists  when one adds U(1) gauge fields. This should alleviate some technical issues in the path of constructing smooth geometries in the F1-NS5-P frame.

Finally, in our analysis, we focused only on momentum-carrying modes that preserve the isometry of  the $T^4$.
It would be interesting to consider momentum-carrying waves coming from  fluctuations of branes along some of the torus directions, and which break this isometry. These fluctuations give rise to U(1) vector fields even in the D1-D5-P duality frame. Furthermore, one can obtain examples of such solutions  by performing a 9-11 flip on our solutions with D0-D4 density modes. Thus, the solutions we have constructed provide a simple way to access dynamics of  compactification tori,  while also preserving the isotropy of the $T^4$. 
We therefore expect the D0-D4 fluctuations to provide qualitatively similar results to analyzing more complicated excitations on the $T^4$ of IIA or IIB supergravity \cite{Kanitscheider:2007wq, Bakhshaei:2018vux, Heidmann:2021cms}.

\section*{Acknowledgements}

We would like to thank Davide Bufalini, Soumangsu Chakraborty, Pierre Heidmann, Anthony Houppe, Bogdan Ganchev, Monica Guica, Nicolas Kovensky, and David Turton for interesting discussions.
This work is supported in part by the ANR grant Black-dS-String ANR-16-CE31-0004-01, by the John Templeton Foundation grant 61149, by the ERC Grants 787320-QBH Structure and 772408-Stringlandscape, and by the DOE grant DE-SC0011687.

\appendix

\section{Chain of dualities}
\label{app:Dualitites}

In this appendix we present the explicit solutions for the intermediate steps in the two duality chains that we discussed in section~\ref{sec:Construction}.
In the first part we present the steps \eqref{eq:DualChain} that lead from the F1-P system with a non-trivial profile  \eqref{eq:F1PSol} to the NS5-P system with local D0-D4 charges \eqref{eq:NS5P-D0D4}.
In the second subsection we then present the chain of dualities \eqref{eq:DualChain2} which is used to write the latter solution in the D1-D5 frame of \cite{Giusto:2013rxa}.
This allows us to consistently add a D1-brane charge which corresponds to adding an F1 charge in the F1-NS5 frame. 

\subsection{Generating the NS5-P-(D0-D4) solution}

\subsubsection*{F1-P}

The starting point is the F1-P configuration in Type IIB theory depicted in figure~\ref{fig:F1P}: Take the fundamental string to wrap the $S^1(y)$ circle and have a non-trivial profile $F(v)$ along one of the directions of the $T^4$, which we call $z_9$. 
Add momentum along the $y$ direction and distribute (smear) the string charge along the four-torus while keeping all the charge localized at a point in $\mathbb{R}^4$.
The supergravity solution corresponding to such a configuration is given by \cite{Callan:1995hn, Dabholkar:1995nc}
\begin{subequations}
	\label{eq:AStep1}
	\begin{align}
		ds^2 &= -\frac2{H_5} \, dv\left[ du -\frac{\dot F^2(v)}{2}\,  \left(H_5- 1\right) \, dv + \dot F(v) \, \left(H_5- 1\right) \, dz^9\right] +  dx^i\, dx^i+ dz^a\, dz^a\,,\\*
		B &= -\left(1- \frac{1}{H_5}\right)\,\left[ du \wedge dv + \dot F(v)\, dv \wedge dz^9\right]\,,\qquad 
		e^{2 \phi} =\frac{1}{ H_5}\,,
	\end{align}
\end{subequations}
with all other fields vanishing.
In the above, $u$ and $v$ are null coordinates \eqref{eq:NullCC} and $H_5$ is a harmonic function associated with the F1-string and is given by \eqref{eq:HarFun}.

\subsubsection*{S-duality to D1-P}

The next step is to perform an S-duality \eqref{eq:SDuality} which yields
\begin{subequations}
	\label{eq:AStep2}
	\begin{align}
		ds^2 &= - \frac{2}{\sqrt{H_5}} \, dv\left( du + \frac{\dot F^2(v)}{2}\, \left(1 - H_5\right)\, dv - \dot F(v)\, \left(1 - H_5\right)\, \dot F(v)\, dz^9 \right)\nonumber\\*
		&\quad  + \sqrt{H_5} \left( dx^i\, dx^i+ dz^a\, dz^a\right)\,,\\
		B&=0\,, \qquad e^{2 \phi} = H_5\,, \\
		C_0 & = 0 \,,\\
		C_2 &= \left(1- \frac{1}{H_5}\right)\, du\wedge dv +  \dot F(v)\left(1- \frac{1}{H_5}\right)\, dv\wedge dz^9\,,\\
		C_4 & = 0 \,,\\
		C_6 &= \gamma \wedge \left( \dot F(v)\, dv \wedge dz^6 \wedge dz^7 \wedge dz^8 + \widehat{\rm vol}_4\right)\,,
	\end{align}
\end{subequations}
where we have introduced a two-form $\gamma$ such that
\begin{align}
	d \gamma \equiv *_4 dH\,,
\end{align}
and used the volume form of the $T^4$
\begin{align}
	\label{eq:AVolForm}
	\widehat{\rm vol}_4 \equiv dz^6 \wedge dz^7 \wedge dz^8\wedge dz^9\,.
\end{align}
This solution describes a D1-brane wrapping the $S^1(y)$ circle and carrying momentum along that direction. 
The D1-brane is smeared along the $T^4$, with a non-trivial profile along the $z_9$, while being located at the origin or the base space. 
We use the democratic formalism (see Appendix~\ref{app:Conventions}), which we have used to determine $C_6$ by imposing the duality condition between $F_3$ and $F_7$.%

\subsubsection*{T-dual along $z_9$ to D2-P with local D0-F1 charges}

Next we perform T-dualities \eqref{eq:TDuality} along all four directions of the torus, and we begin with the ``special'' direction $z_9$. 
When performing this duality, following Figure~\ref{fig:LocalFig}, the decomposition of the local charges into those along the $y$ and the $z_9$ direction become important. 
The result is a configuration in Type IIA theory: a D2-brane (wrapping the $y$ and $z_9$ directions) with a momentum along $y$, on which we find D0 and F1 charges (the latter wrapping the $z_9$ direction), which have varying densities along the $y$ direction.
The corresponding supergravity solution is
\begin{subequations}
	\label{eq:AStep3}
	\begin{align}
		ds^2& = - \frac{2}{\sqrt H_5} \, dv\,\left[ du - \frac{\dot F^2(v)}{2}\left(1- \frac{1}{H_5}\right)\, dv\right] + \sqrt{H_5}\left( dx^i \, dx^i + \sum_{a = 6}^8 dz^a \,dz^a\right)\nonumber\\*   &\quad+  \frac1{\sqrt{H_5}}\left(dz^9\right)^2\,,\\
		B_2 &=  \dot F(v) \left(1- \frac{1}{H_5}\right) dv \wedge dz^9\,,\qquad e^{2 \phi} = \sqrt{H_5}\,. \\
		C_1 &=  \dot F(v) \left(1- \frac{1}{H_5}\right)dv\,,\\
		C_3 &= \left(1- \frac{1}{H_5}\right)\, du \wedge dv\wedge dz^9 \,,\\
		C_5 &= \gamma \wedge dz^6 \wedge dz^7 \wedge dz^8\,,\\
		C_7 &= \frac{\dot F(v)}{H_5}\, \gamma \wedge dv\wedge \widehat{\rm vol}_4 \,.
	\end{align}
\end{subequations}
 In the above solution, the $y$-, or more appropriately $v$-, dependent distribution of D0 and F1 charges is seen in the dependence on $\dot F(v)$ that appears in $B_2$, which is sourced by fundamental strings, and $C_1$ ($C_7$) which is electrically (magnetically) sourced by D0-branes.  
 On the other hand, $C_3$ and $C_5$, which are sourced by D2-branes, are independent of $\dot F(v)$.

\subsubsection*{T-dualities along $z_8$, $z_7$ and $z_6$ to the D5-P with local D3-F1 charges}

The three T-dualities along $z_8$, $z_7$, and $z_6$ (in that order) are
very similar and thus we perform them together. 
The final result is a configuration in Type IIB theory where the D2-brane now becomes a D5-brane wrapping the $S^1(y)$ circle and all four directions of the $T^4$, while the $\dot F(v)$ dependent fields are now sourced by local D3 and F1 charges:
\begin{subequations}
	\label{eq:AStep4}
	\begin{align}
		ds^2& = - \frac{2}{\sqrt H_5} \, dv\,\left[ du - \frac{\dot F^2(v)}{2}\left(1- \frac{1}{H_5}\right)\, dv\right] + \sqrt{H_5} dx^i \, dx^i +\frac{1}{\sqrt H_5}\, dz^a\, dz^a\,,\\
		B_2 &=  \dot F(v) \left(1- \frac{1}{H_5}\right) dv \wedge dz^9\,,\qquad e^{2\phi } = \frac{1}{H_5}\,,\\
		C_0 &=  0\,,\\
		C_2 &= \gamma \,,\\
		C_4 &= - \frac{\dot F(v)}{H_5}\, \gamma \wedge dv\wedge dz^9 -\dot F(v) \left(1- \frac{1}{H_5}\right) dv\wedge dz^6 \wedge dz^7 \wedge dz^8 \,,\\
		C_6 &= \left(1- \frac{1}{H_5}\right)du\wedge dv\wedge \widehat {\rm vol}_4\,,\\
		C_8 &= 0\,.
	\end{align}
\end{subequations}

\subsubsection*{S-duality to NS5-P with local D3-D1 charges}
Since our aim is to obtain a solution corresponding to a configuration with NS5-P charges, we continue with another S-duality.
Essentially, this  only exchanges the D5-brane for an NS5-brane and the D1 local charges with F1 charge distribution:
\begin{subequations}
	\label{eq:AStep5}
	\begin{align}
		ds^2 &= -2 dv \left[ du - \frac{\dot F^2(v)}{2}\left(1- \frac{1}{H_5}\right)\, dv\right]+ H_5 \, dx^i \, dx^i + \, dz^a \, dz^a\,,\\
		B_2 & = \gamma\,, \qquad e^{2 \phi} = H_5\,,\\
		C_0 & = 0\,,\\
		C_2 &= - \dot F(v) \left(1- \frac{1}{H_5}\right)\, dv\wedge dz^9\,,\\
		C_4 &= - \dot F(v) \, \gamma \wedge dv \wedge dz^9 - \dot F(v) \left(1- \frac{1}{H_5}\right)\, dv\wedge dz^6 \wedge dz^7\wedge dz^8\,,\\
		C_6 &= - \dot F(v) \, \gamma \wedge dv \wedge dz^6 \wedge dz^7 \wedge dz^8\,,\\
		C_8 & = 0\,.
	\end{align}
\end{subequations}

\subsubsection*{T-duality  to NS5-P with local D0-D4 charges}
Finally, we perform another T-duality along $z_9$, which lands us in the desired configuration: an NS5-brane with momentum along the $y$-direction with  D0- and D4-brane charges which vary along the $S^1(y)$ circle
\begin{subequations}
	\label{eq:AStep6}
	\begin{align}
		ds^2 &= - 2 dv\, \left[ du - \frac{\dot F(v)^2}{2}\,\left(1- \frac1{H_5}\right) \,dv\right] + H \, dx^i\, dx^i+ dz^a \, dz^a\,,\\
		B_2 &= \gamma\,, \qquad e^{2\phi} = {H_5}\,,\\
		C_1 &= - \dot F(v)\left(1- \frac1{H_5}\right) \, dv\,, \\
		C_3 &= - \dot F(v)\, \gamma \wedge dv\,,\\
		C_5 &=- \dot F(v)\left(1- \frac1{H_5}\right) \, dv\wedge \widehat {\rm vol}_4 = C_1 \wedge \widehat {\rm vol}_4\,,\\
		C_7 & = - \dot F(v)\, \gamma \wedge dv \wedge 	\widehat {\rm vol}_4 = C_3 \wedge \widehat {\rm vol}_4\,,
	\end{align}
\end{subequations}
which is the solution \eqref{eq:NS5P-D0D4} presented in the main text. 
Unlike any of the previous solutions presented in this appendix, \eqref{eq:AStep6} depends on the $T^4$ only through its volume form \eqref{eq:AVolForm}.

\subsection{Adding F1 charge by using a Gibbons-Hawking base}
\label{sec:Tpsi-duality}

The solution \eqref{eq:AStep6} (or equivalently \eqref{eq:NS5P-D0D4} of the main text) is asymptotically a two-charge solution. 
To make contact with the microstate geometries programme,  we would like to construct a solution which has three charges.
We choose to add to the configuration an additional fundamental string that wraps the $S^1(y)$ circle and is smeared along the $T^4$. 

We do so in a roundabout way: We write the four-dimensional flat metric in Gibbons-Hawking form and T-dualize along the Gibbons-Hawking fiber.
If we then perform an S-duality, the resulting configuration should be described in terms of the complete ansatz for the D1-D5 system constructed in \cite{Giusto:2013rxa}.
Adding a source corresponding to a D1-brane in this duality frame is equivalent to adding a fundamental string in the NS5-P frame, only that in the former frame we know all fields which get excited as a consequence of adding a new object into the configuration.

Begin by writing the flat base space metric in \eqref{eq:AStep6} as%
\footnote{
\label{footnote:Spherical}
In what follows we do not specify the coordinates used in the Gibbons-Hawking ansatz.
However, one can introduce spherical coordinates for $\mathbb{R}^4$ whose metric can be written as
\begin{align*}
		ds_4^2 &= dr^2 + r^2\left(d\theta^2 + \sin^2\theta\, d\varphi_1^2 + \cos^2\theta\, d\varphi_2^2\right).
\end{align*}
To rewrite this metric in the Gibbons-Hawking form, we introduce new coordinates as $r \equiv 2\sqrt{\rho}$, $ \tilde\theta \equiv 2 \theta$, $\psi \equiv \varphi_1 + \varphi_2$, and $\phi \equiv \varphi_2 - \varphi_1$, 
where the ranges of various coordinates are taken to be $\varphi_{1,2} \in [0, 2\pi)$, $\psi \in [0, 4\pi)$, and $\phi \in [0, 2\pi)$, while $r$ and $\rho$ are both taken to be non-negative.
The metric becomes
\begin{align}
	ds_4^2= \rho \, (d\psi + \cos \tilde \theta\, d\phi)^2 + \frac{1}{\rho}\left( d\rho^2 + \rho^2\left(d\tilde \theta^2 + \sin^2\tilde\theta \, d\phi^2\right)\right) \,,
\end{align}
and one can read off that $V = \rho^{-1}$ and $  A= \cos\tilde\theta \, d\phi$. Furthermore $H_5 = 1 + \frac{Q_5}{4\rho}$, and is thus harmonic even in $\mathbb{R}^3$.
}
\begin{align}
	\label{eq:GHForm2}
	dx^i\,dx^i = \frac{1}{V}(d\psi + A)^2 + V\, ds_3^2\,,
\end{align}
where $ds_3^2$ denotes the flat metric on $\mathbb{R}^3$.
Recall  that we need to impose the following constraints on the function $V$ and one-form $A$
\begin{align}
	*_3 dA = dV\,,\quad \Longrightarrow \quad 
	*d*dV=0\,, \qquad d*dA = 0\,,
\end{align}
which also means that the warp factor, $V$, is a harmonic function in  $\mathbb{R}^3$. 
The metric \eqref{eq:GHForm2} is invariant under a simultaneous rescaling of the coordinates, the function  $V$, and one-form $A$, which we can fix by setting the periodicity of $\psi$ to be $4\pi$.

Now assume that $\psi$ denotes an isometry direction of the solution. 
Then one can decompose 
\begin{align}
	*_4 dH_5 = (d\psi + A) \wedge*_3 d H_5\,,
\end{align}
and%
\footnote{For example, in spherical coordinates (see footnote~\ref{footnote:Spherical})  $\gh = \frac14\,Q_5\, \cos \tilde \theta\, d\phi$ and $\ghh = 0$.}
\begin{align}
	\gamma \equiv -\left(d\psi + A\right) \wedge \gh + \ghh\,,
\end{align}
where the one-form $\gh$ and the two-form $\ghh$ are determined from the definition \eqref{eq:dgamma} by
\begin{align}
	\label{eq:Gamma12eq}
	d\gh = *_3 dH_5 \,, \qquad d\ghh = *_3 dV \wedge \gh\,.
\end{align}

\subsubsection*{T-duality along the Gibbons-Hawking fiber}

We now use the T-duality rules to dualize along the Gibbons-Hawking fiber $\psi$.
However, after performing the transformation, we need to change the sign of $\psi$
\begin{align}
	\label{eq:PsiFliP}
	\psi \to - \psi\,,
\end{align}
to obtain 
\begin{subequations}
	\label{eq:AStep7}
	\begin{align}
		ds^2 &= - 2 \,dv \left[ du - \frac{\dot F^2(v)}{2}\left(1 - \frac{1}{H_5}\right) dv\right] + V\left[\frac{1}{H_5}\left(d\psi+ \gh\right)^2 + H_5 \,ds_3^2 \right] + ds_4^2\,,\\
		B_2 & = A\wedge d\psi+ \ghh\,,\qquad e^{2\phi} = V\,,\\
		C_0 & = 0\,,\\
		C_2 & = \dot F(v) \left(1- \frac{1}{H_5}\right)\, dv\wedge(d\psi + \gh)- \dot F \,dv \wedge \gh\,,\\
		C_4 &= \dot F(v)\, dv\wedge \left(d\psi + \gh\right) \wedge \ghh \,,
		%
	\end{align}
\end{subequations} 
where the sign flip \eqref{eq:PsiFliP} ensures that the first equation of \eqref{eq:Gamma12eq} now serves as the constraint between the one-form and scalar function in the new Gibbons-Hawking base-space metric.

\subsubsection*{S-duality to the D1-D5 frame}

S-dualizing the above solution puts us in the D1-D5 frame, and the resulting configuration fits within the ansatz of \cite{Giusto:2013rxa}.
In this transformation, and only in this transformation alone, we choose $b=-c=-1$ when performing the S-duality \eqref{eq:SDuality}.
This allows us to compare the resulting solution with the complete ansatz of   \cite{Giusto:2013rxa} without changing the signs of the fields and furthermore, when transforming back to the NS5-P system we can take $b = -c = 1$ which is the inverse transformation.
We find
\begin{subequations}
	\label{eq:AStep8}
	\begin{align}
		ds^2 & = - \frac{2}{\sqrt{V}} \, dv\,\left[ du  - \frac{\dot F^2(v)}{2}\left(1 - \frac{1}{H_5}\right) dv\right]+ \sqrt V \left[ \frac{1}{H_5}\left(d\psi + \gh\right)^2 + H_5\, ds_3^2\right]\nonumber\\*
		& \quad + \frac{1}{\sqrt{V}}\, d\hat s_4^2\,,\\
		B_2 &= \dot F(v) \left[\left(1-\frac{1}{H_5}\right)\, \left(d\psi + \gh\right)- \gh\right]\wedge dv\,,\qquad e^{2\phi}  = \frac{1}{V}\,,\\
		C_0 & = 0\,,\\
		C_2 &= A \wedge \left(d\psi + \gh\right) + \ghh- A\wedge \gh\,,\\
		C_4 &= - \dot F(v)\, \left[\frac1{H_5} \,\left(d\psi + \gh\right) \wedge \left(\ghh - A\wedge \gh\right) + \gh \wedge \ghh\right]\wedge dv\,.
		%
		%
		%
	\end{align}
\end{subequations}
At this point one can recombine the Gibbons-Hawking decomposition of the base space (including the forms), compare the solution \eqref{eq:AStep8} with the complete ansatz of \cite{Giusto:2013rxa} and read off the ansatz quantities,%
\footnote{Once this is done, one can check that the read-off quantities solve the BPS equations \cite{Bena:2011dd, Giusto:2013rxa}.}
however, this is not central to our analysis.

\subsubsection*{Adding a D1 charge}

What is important for us is that the harmonic function corresponding to D1-brane sources is precisely known in the complete ansatz \cite{Giusto:2013rxa}.%
\footnote{In the notation commonly used in the microstate geometries literature dealing with the D1-D5 system \cite{Bena:2011dd, Giusto:2013bda, Bena:2015bea, Bena:2017xbt} (see also appendix E.7 of \cite{Giusto:2013rxa}), this is the scalar function $Z_1$. 
Note that in addition one would need to turn on a contribution to the gauge field $C_6$, which would ensure, in the democratic formalism, appropriate self-duality properties of the gauge field strengths.
However, we will  determine higher-order gauge fields only after the last duality transformation.}
Thus denoting this harmonic function with $H_1$ (see \eqref{eq:HarFun2}), we find that the new solution is given by
\begin{subequations}
	\label{eq:AStep9}
	\begin{align}
		ds^2 & = - \frac{2}{\sqrt{V\, H_1}} \, dv\,\left[ du  - \frac{\dot F^2(v)}{2}\left(1 - \frac{1}{H_5}\right) dv\right]+ \sqrt {V\, H_1}\, \left[ \frac{1}{H_5}\left(d\psi + \gh\right)^2 + H_5\, ds_3^2\right]\nonumber\\&
		\quad + \sqrt{\frac{H_1}{V}}\, d\hat s_4^2\,,\\
		B_2 &= \dot F(v) \left[\left(1-\frac{1}{H_5}\right)\, \left(d\psi + \gh\right)- \gh\right]\wedge dv\,,\qquad e^{2\phi} = \frac{H_1}{V}\,,\\
		C_0 & = 0\,,\\
		C_2 &= -\frac{1}{H_1}du\wedge dv+A \wedge \left(d\psi + \gh\right) + \ghh- A\wedge \gh\,,\\
		C_4 &= - \dot F(v)\, \left[\frac1{H_5} \,\left(d\psi + \gh\right) \wedge \left(\ghh - A\wedge \gh\right) + \gh \wedge \ghh\right]\wedge dv\,.
	\end{align}
\end{subequations}
It is straightforward to check that this supersymmetric torus-independent D1-D5-frame solution (\ref{eq:AStep9}) solves the equations governing all such solutions \cite{Giusto:2013rxa}.

\subsubsection*{S-dual to F1-NS5 frame in Type IIB}
To return to the NS5-P system, we need to first perform an S-duality and then a T-duality along $\psi$.
Using $b = -c = 1$, which ensures that this is the inverse transformation of the one used to arrive at \eqref{eq:AStep8}, we obtain
\begin{subequations}
	\label{eq:AStep10}
	\begin{align}
		ds^2 &= -  \frac{2}{H_1}\, dv \left[ du - \frac{\dot F^2(v)}{2}\left(1 - \frac{1}{H_5}\right) dv\right] + V\left[\frac{1}{H_5}\left(d\psi+ \gh\right)^2 + H_5\, ds_3^2 \right] + ds_4^2\,,\\
		B_2 & = - \frac{1}{H_1}du\wedge dv+A\wedge d\psi+ \ghh\,,\qquad e^{2\phi} = \frac{V}{H_1}\,,\\
		C_0 & = 0\,,\\
		C_2 & = -\dot F(v) \left[\left(1-\frac{1}{H_5}\right)\, \left(d\psi + \gh\right)- \gh\right]\wedge dv\,,\\
		C_4 &= -\dot F(v)\, \left(d\psi + \gh\right) \wedge \ghh \wedge dv\,.
	\end{align}
\end{subequations} 

\subsubsection*{T-dual to the F1-NS5 system in Type IIA}
To return to the original system we perform a final T-duality along the $
\psi$ direction, which has to be again followed by a sign flip \eqref{eq:PsiFliP}.
Furthermore, in order to compare the final solution to the two-charge case \eqref{eq:AStep6}, we also exchange
$\dot F(v) \to - \dot F(v)$.
Then one finds
\begin{subequations}
	\label{eq:AStep11}
	\begin{align}
		ds^2 &= - \frac{2}{H_1} dv\, \left[ du - \frac{\dot F(v)^2}{2}\,\left(1- \frac1{H_5}\right) \,dv\right] + H_5 \, dx^i\, dx^i+ dz^a \, dz^a\,,\\*
		B_2 &= - \frac{1}{H_1}\, du \wedge dv+ \gamma\,, \qquad e^{2\phi} = \frac{H_5}{H_1}\,,\\*
		C_1 &=-  \dot F(v)\left(1- \frac1{H_5}\right) \, dv\,, \\*
		C_3 &=  -\dot F(v)\, \gamma \wedge dv\,,
	\end{align}
\end{subequations}
where we have recombined the decompositions along the Gibbons-Hawking fiber.
After the remaining RR gauge fields are computed, this solution matches the one presented in the main text in Equation \eqref{eq:NS5F1P-D0D4}.

\section{Conventions}
\label{app:Conventions}

\subsection*{Democratic  formalism}

When dealing with brane sources it is useful to introduce the  democratic formalism \cite{Bergshoeff:2001pv} which effectively doubles the number of gauge fields in the theory, but introduces self-duality constraints on the field strengths so that the number of degrees of freedom remains unchanged.
This democracy is imposed only on the Ramond-Ramond gauge fields $C_p$, while we keep only one NS-NS gauge field $B$, with a three-form field strength 
\begin{align}
	H_3 = dB\,.
\end{align}
The RR field strengths are defined as
\begin{align}
	\label{eq:DemFormFielStre}
	F_p \equiv dC_{p-1}- H_3 \wedge C_{p-3}\,,
\end{align}
which satisfy modified Bianchi identities $dF_p = H_3 \wedge F_{p-2}$.

In each of the Type II theories, we introduce additional RR gauge field potentials, so that  for Type IIA we consider $\{C_1, C_3, C_5, C_7\}$ and $\{C_0, C_2, C_4, C_6, C_8\}$ for Type IIB.
However, the number of degrees of freedom is kept constant by imposing 
\begin{subequations}
	\label{eq:DemFormSD}
	\begin{align}
		&(IIA):&&  F_2 = * F_8 \,, \quad F_4 = - * F_6\,, \quad F_6 = * F_4\,, \quad F_8 = - * F_2\,,\\*
		&(IIB):&& F_1=*F_9,\quad F_3=-*F_7, \quad F_5=*F_5,\quad F_7=-*F_3,\quad F_9=*F_1\,, 
	\end{align}
\end{subequations}
which imply that the field strengths $F_p$ and $F_{10-p}$ essentially convey the same information. 
Note that we follow the conventions of \cite{Giusto:2013rxa}, where the Hodge dual of a $k$-form in a $D$-dimensional spacetime is given by
\begin{align}
	*X_k  \equiv \frac{1}{k! (D-k)!}\, \epsilon_{m_1\ldots m_{D-k}, n_{D-k+1} \ldots n_D}\, X^{n_{D-k+1} \ldots n_D}\, e^{m_1}\wedge \ldots e^{m_{D-k}}\,.
\end{align}
Furthermore, we choose the orientation
\begin{align}
	 \epsilon^{+-12346789}  =\epsilon^{1234} = 1\,.
\end{align}

\subsection*{S-duality}

Define a complex field as a combination of the axion field and the dilaton and combine the two-form gauge potentials into a vector
\begin{align}
	\lambda \equiv C_0 + i \, e^{-\phi}\,,\qquad 	T = \begin{pmatrix}
		B_2 \\ C_2
	\end{pmatrix}\,.
\end{align}
Type IIB theories are invariant under a transformation generated by $U \in SL(2, \mathbb{R})$ 
\begin{align}
	U = \begin{pmatrix}
		a& b \\ c & d
	\end{pmatrix}\,, \quad\text{with} \quad  a\,d - b\,c  = 1\,,
\end{align}
such that 
\begin{align}
	\label{eq:SDuality}
	\lambda \to\tilde \lambda =  \frac{a \lambda + b}{c \lambda + d} \,, \qquad T \to\tilde T =  U \, T\,,
\end{align}
while the five-form gauge field strength, $F_5$, and the ten-dimensional metric in the Einstein frame are invariant.

In the main text we consider only a  $\mathbb{Z}_2$ subgroup of $SL(2, \mathbb{R})$  transformations where 
\begin{align}
	\label{eq:SDualZ2}
	a = d = 0\,, \qquad b = -c = \pm 1.
\end{align} 
Unless explicitly stated otherwise, we choose $b = -c = 1$ whenever we perform an S-duality transformation.
In addition, in all of the solutions considered, the axion field $C_0$ is vanishing.
Then the effect of such a transformation, with either choice of sign for $b$ and $c$, results in the inversion of the dilaton field 
\begin{align}
	\tilde \phi = - \phi\,,
\end{align}
and the following change of the metric in the string frame 
\begin{align}
	\tilde G_{\mu\nu} & =  e^{-\phi}\, G_{\mu\nu}\,.
\end{align}
Furthermore, the two-form gauge fields are interchanged up to a minus sign
\begin{align}
	\tilde B_2 &= \pm C_2\,, & \tilde C_2 &= \mp B_2\,,
\end{align}
where the upper (lower) sign corresponds to $b = +1$ ($b=-1$).
For either sign, the invariance of $F_5$ implies that the four-form gauge field transforms as
\begin{align}
	\tilde C_4 & = C_4 - B_2 \wedge C_2 \,.
\end{align}
Higher-form gauge fields can be calculated by using the duality rules of the democratic formalism \eqref{eq:DemFormSD} and \eqref{eq:DemFormFielStre}. 
The effect of this particular transformation is thus to effectively exchange the two-form gauge potentials.

\subsection*{T-duality}

For performing T-duality transformations we use the conventions of \cite{DallAgata:2010srl}, which are convenient when one works in the democratic formalism.
Assume that we are performing a T-duality along an isometry direction coordinatized  by $y$. 
Rewrite the initial string frame metric and gauge fields as
\begin{subequations}
	\label{eq:TDuality}
	\begin{align}
		ds^2 &= G_{yy}\left( dy + A_\mu \, dx^\mu\right)^2 + \widehat{g}_{\mu \nu} \, dx^\mu \, dx^\nu \\
		B_2 &= B_{\mu y} dx^\mu\wedge \left( dy + A_\mu \, dx^\mu\right) + \widehat B_2\,,\\
		C_p &= C_{p-1}^y\wedge \left( dy + A_\mu \, dx^\mu\right) + \widehat C_p\,,
	\end{align}
\end{subequations}
where the forms $\widehat B_2$, $\widehat C_p$ and $\widehat C_{p-1}^y$ do not have any legs along $y$.
After applying the rules of a T-duality transformation \cite{Buscher:1987qj, Buscher:1987sk}, the new fields (denoted with the tilde)  are
\begin{subequations}
	\begin{align}
		d\widetilde s^2 &= G^{-1}_{yy}\left( dy - B_{\mu y} \, dx^\mu\right)^2 + \widehat{g}_{\mu \nu} \, dx^\mu \, dx^\nu \\
		\widetilde B_2 &= -A_{\mu} dx^\mu\wedge dy + \widehat B_2\,,\\
		\widetilde C_p &= \widehat C_{p-1} \wedge \left( dy -B_{\mu y} \, dx^\mu\right) +  C^y_p\,,\\
		e^{2 \widetilde \phi} &= G_{yy}^{-1}\, e^{2\phi}\,.
	\end{align}
\end{subequations}


\bibliographystyle{JHEP}

\bibliography{ato0}

\end{document}